\documentclass[fleqn,usenatbib]{mnras}
\usepackage[cp1251]{inputenc}
\usepackage[T1]{fontenc} 
\usepackage{aecompl}
\usepackage{graphicx,epsfig}
\usepackage{floatflt}
\usepackage {amssymb}
\usepackage {amsmath}
\usepackage {longtable}
\usepackage[usenames]{color}
\title[Structure of orbits in ring systems]
    {The structure and stability of orbits in Hoag-like ring systems}
\author[Bannikova]
  {Elena Yu.~Bannikova$^{1,2}$ \thanks {E-mail: bannikova@astron.kharkov.ua}\\
   $^1$   Institute of Radio Astronomy of Nat.Ac.Sci. of Ukraine, Mystetstv 4,
   61022 Kharkiv, Ukraine \\
   $^2$  V.N.Karazin Kharkiv National University,
   Svobody Sq. 4, 61022 Kharkiv, Ukraine 
   }

\date{Accepted  .  Received ; in
original form }

\pagerange{\pageref{firstpage}--\pageref{lastpage}} \pubyear{2017}

\def\LaTeX{L\kern-.36em\raise.3ex\hbox{a}\kern-.15em
    T\kern-.1667em\lower.7ex\hbox{E}\kern-.125emX}

\begin{document}

\label{firstpage}

\maketitle

\begin{abstract}
Ring galaxies are amazing objects exemplified by the famous case of the  Hoag's Object. Here the mass of the central galaxy may be comparable to the mass of the ring, making it a difficult case to model mechanically. In a previous paper, it was shown that the outer potential of a torus (ring) can be represented with good accuracy by the potential of a massive circle with the same mass. This approach allows us to simplify the problem of the particle motion in the gravitational field of a torus associated with a central mass by replacing the torus with a massive circle. In such a system there is a circle of unstable equilibrium that we call "Lagrangian circle" (LC). Stable circular orbits exist only in some region limited by the last possible circular orbit related to the disappearance of the extrema of the effective potential. We call this orbit "the outermost stable circular orbit" (OSCO) by analogy with the innermost stable circular orbit (ISCO) in the relativistic case of a black hole. Under these conditions, there is a region between OSCO and LC where the circular motion is not possible due to the competition between the gravitational forces by the central mass and the ring. As a result, a gap in the matter distribution can form in Hoag-like system with massive rings.
\end{abstract}


\section{INTRODUCTION}
\label{sec:intro}
There are astrophysical objects in which gravitating toroidal (ring) structures are observed. 
Examples are the obscuring tori in active galactic nuclei \citep{Antonucci1993, Jaffe2004},  
ring galaxies, and polar ring galaxies \citep{Whitmore1990, Moiseev2011, Moiseev2015}. Ring distribution of dark matter was discovered by means of gravitational lensing effects in the galaxy cluster Cl 0024+17 \citep{Jee2007}. In objects such as ring galaxies, the mass of the ring is comparable with the central massive core. This is the case of the Hoag's Object, where a central galaxy is surrounded by a massive ring with on-going star formation \citep{Hoag1950}. Optical \citep{Schweizer1987, Finkelman2011} and radio \citep{Brosch2013} observations show that the mass of the ring is approximately $1/3$ of the central galaxy. Observations have identified sources which appear to be similar to Hoag's Object \citep{Moiseev2011, FinkelmanBrosch2011, MutluPakdil2017}. Therefore, it becomes interesting to investigate the influence of the outer gravitational field of the massive ring on the distribution of matter.

From the geometrical point of view, a ring is a particular case of a torus with a minor radius much smaller than the major one (a thin torus). It is known that the potential of the torus is quite cumbersome and uneasy to handle even in the case of a homogeneous circular structure. To overcome this problem, by applying general considerations the outer torus potential has been replaced by that of an infinitely thin ring \citep{Woodward1992}, since the torus potential tends to that of an infinitely thin ring far from the torus surface. In other cases, the torus potential has been approximated by the potential of a thick disc \citep{Hure2009}. Recent advances in the investigation of the torus potential \citep{Bannikova2011, Fukushima2016} can simplify further the problem of the particle motion in this gravitational field. \citet{Bannikova2011} have shown that the outer potential of a homogeneous circular torus is quite accurately represented by that of a massive circle (MC) of the same mass and of a radius equal to the major radius of the torus. The representation of the outer potential of the torus by the MC potential works well at any point up to the torus surface with small errors depending on the thickness of the torus cross-section. In this study we analyze the dynamics of a particle in the gravitational field of a massive ring with a central core.

The paper is organized as follows. The idea about Lagrangian circle is presented in Section~\ref{sec:LC}. In Section~\ref{sec:OSCO} we consider the features of a particle motion in the equatorial plane, and in Section~\ref{sec:merid} the spatial cases of closed and regular particle orbits in the meridional plane. A possible application of these results to Hoag's Object is discussed in Section~\ref{sec:discussion}.

\section{LAGRANGIAN CIRCLE}
\label{sec:LC}

Consider the motion of a test particle in the outer \footnote{Here, the adjective "outer" applies to any point outside the torus volume where the matter density is null.} gravitational potential of a non-rotating homogeneous circular torus and a central mass (core). The gravitational forces from the central mass and from the torus will act in opposite directions, so they must equilibrate at some distance in the symmetry plane. By some kind of analogy with the Lagrangian point $L_1$ in the three-body problem, we call "Lagrangian circle" (LC) the geometrical locus where the balance of these forces is realized. LC is obviously a place of unstable equilibrium. Small perturbations determine the destiny of a particle, which will be captured either by the central mass or by the torus. In order to find the LC radius, we need to obtain the gravitational force of the torus $F_{torus,\rho}=\partial \varphi_{torus}/ \partial \rho$.

\begin{figure}
\includegraphics[width = 80mm]{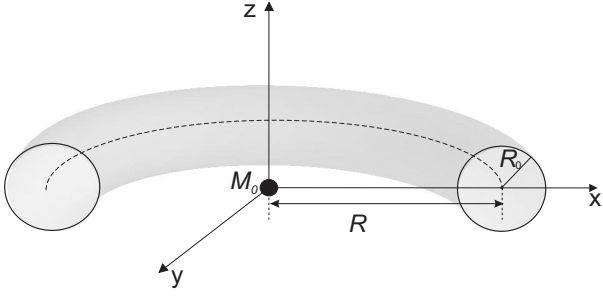}
\caption{Scheme of a torus with a central mass. The massive circle (MC) is indicated by the dashed curve.}
\label{Fig:1}
\end{figure}

The circular torus is characterized by two geometrical parameters: the minor, $R_0$, and major, $R$, radii (Fig.~\ref{Fig:1}). 
We use the dimensionless minor radius (a geometrical parameter) $r_0 = R_0/R$. If the major radius $R=1$, then the parameter $r_0$ determines the geometrical thickness of the torus. The gravitational potential of the homogeneous torus with a circular cross-section \citep{Bannikova2011} is 
\begin{equation}\label{eq:1}
  \varphi_{torus}(\rho,\zeta) = \frac{G M_{mc}}{\pi^2 R
  r_0^2}\int_{-r_0}^{r_0} \int_{-\sqrt{r_0^2 - \eta'^2}}^{\sqrt{r_0^2 - \eta'^2}}
  \phi_r(\rho,\zeta;\eta',\zeta')d\eta'd\zeta',
\end{equation}
where $M_{mc}$ is the torus mass, $\rho = \sqrt{x^2+y^2}/R$, $\zeta = z/R$ are the dimensionless cylindrical coordinates,
$G$ is the gravitational constant;
\begin{equation}\label{eq:2}
  \phi_r(\rho,\zeta;\eta',\zeta') = \sqrt{\frac{(1+\eta')\cdot
  m_r}{\rho}}\cdot K(m_r)
\end{equation}
where $K(m)=\int_0^{\pi/2}d\beta/\sqrt{1-m \sin^2 \beta}$
is the complete elliptic integral of the first kind with its parameter:
\begin{equation}\label{eq:3}
  m_r = \frac{4\rho\cdot (1+\eta')}{(1+\eta'+\rho)^2 +
  (\zeta-\zeta')^2}~.
\end{equation}
According to \citet{Bannikova2011}, the outer potential of the homogeneous circular torus can be represented with a good accuracy by the potential of a massive circle (MC)\footnote{We use the term "massive circle" for an infinitely thin ring having a finite mass. This is the limiting case of the torus with the minor radius tending to zero.} with the same mass and with a radius equal to the torus major radius $R$. Interesting enough, that this result is similar to the known result for a solid sphere: the outer potential of a solid sphere equals exactly that of a massive point. In the considered case, the equivalence of the torus with a MC is not exact: there are small differences in the hole of the torus depending on its geometrical parameter. The good approximate expression for the torus potential for an arbitrary outer point is
\begin{equation}\label{eq:4}
  \varphi_{torus}(\rho,\zeta;r_0) \approx \varphi_{mc}(\rho,\zeta)\cdot \left(
               1 - \frac{r_0^2}{16}+\frac{r_0^2}{16}\frac{\zeta^2 - 1}
                                                         {\zeta^2 + 1}
                          \right),
\end{equation}
where the gravitational potential of a MC with mass $M_{mc}$ and radius $R$ has the form:
\begin{equation}\label{eq:5}
  \varphi_{mc}(\rho,\zeta) = 
  		\frac{G M_{mc}}{\pi R} \cdot \sqrt{\frac{m}{\rho}}\cdot K(m), 
\end{equation}
\begin{equation}\label{eq:6}
m=\frac{4\rho}{(\rho +1)^2+\zeta^2}.
\end{equation}
\begin{figure}
\includegraphics[width = 80mm]{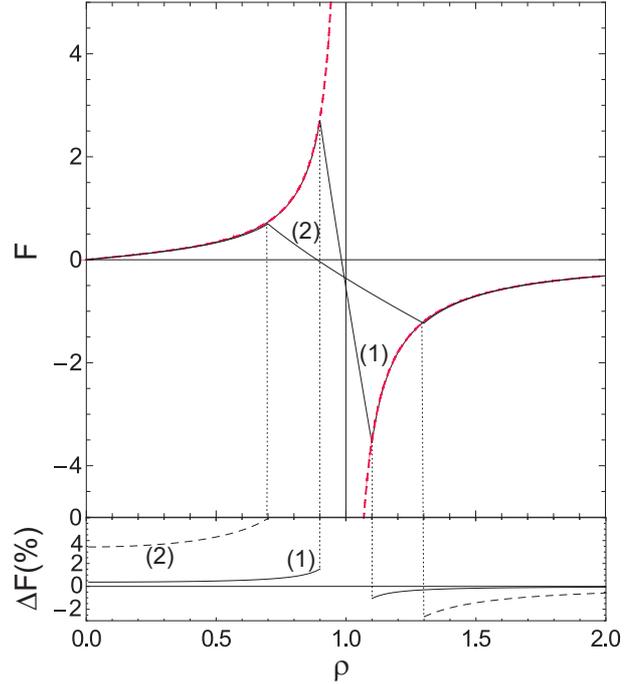}
\caption{{\it Upper panel}: dependence of the radial component of the gravitational force ($\zeta=0$) for a MC with mass $M_{mc}=1$ (a red dashed curve), and for a homogeneous circular torus (solid curves) with the same mass and 1) $r_0=0.1$, 2) $r_0=0.3$. 
{\it Lower panel}: relative residual differences between $F_{torus}$ and $F_{mc}$ (see text). We use the unity system $G=1$, $R=1$, here and for all the following figures.
}
\label{Fig:2}
\end{figure}
For $\zeta =0$, the multiplier in the expression (\ref{eq:4}) is proportional to $r_0^2/8$ and does not depend on the radial distance. It means that we can replace the torus potential by the MC potential $\varphi_{torus} \approx \varphi_{mc}$ for the case of the thin torus (see Fig.~\ref{Fig:2} for the radial component of the gravitational forces): the maximal differences\footnote{In all the following figures we estimate that the differences between a curve and its approximation, $\triangle f = 100\% \times(1-f_{1}/f_{2})$.} $\triangle F=100\%\times(1-F_{torus}/F_{mc})$, are less than $6\%$. The radial component of the force exerted by the MC, $F_{mc,\rho}=\partial \varphi_{mc}/ \partial \rho$ which, after some transformations, is
\[
F_{mc,\rho}=\frac{GM_{mc}}{\pi R^2}\sqrt{\frac{m}{\rho}}\frac{1}{4\rho(1-m)}
	\times 
\]	
\begin{equation}\label{eq:7}	
	\hspace*{1.5cm} \left[
		(2-m(\rho+1))E(m)-  2(1-m)K(m)
	\right],
\end{equation}
where $E(m)=\int_0^{\pi/2}d\beta\sqrt{1-m \sin^2 \beta}$
is the complete elliptical integral of the second kind.
Figure~\ref{Fig:3} shows the dependence of the radial component of the MC force calculated by (\ref{eq:7}), of the central mass force ($F_0=G M_0/\rho^2$), and of their sum for the case $\zeta=0$, $M_0=M_{mc}$ and $R=1$. 
\begin{figure}
\includegraphics[width = 80mm]{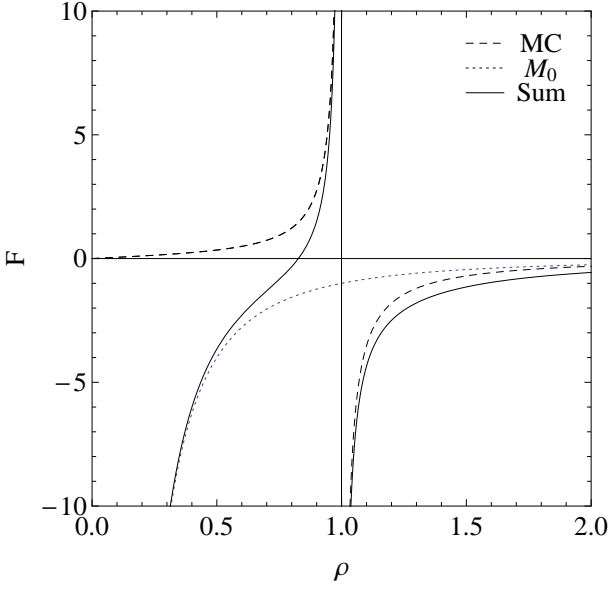}
\caption{Dependence of the radial component of the force from MC (a long dashed curve), from  the central mass (a short dashed curve), and from their sum (a solid curve)  in the MC plane for $M_0=M_{mc}$.}
\label{Fig:3}
\end{figure}
\begin{figure}
\includegraphics[width = 80mm]{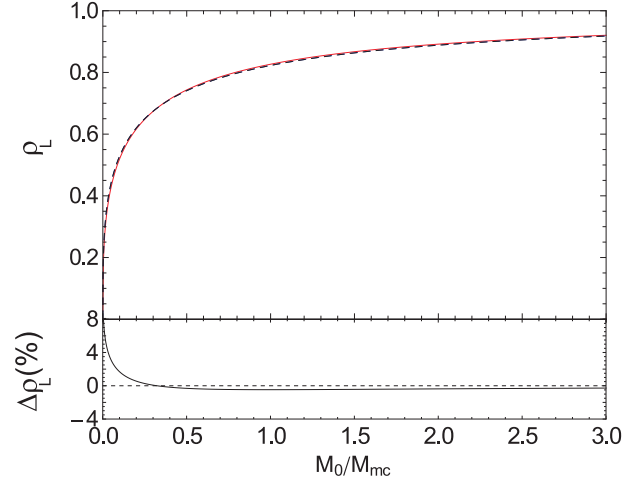}
\caption{{\it Upper panel}: the dependence of the LC radius on the central-to-MC mass ratio is obtained by solving eq.(\ref{eq:13}) numerically (red solid curve) and by the approximated expression (\ref{eq:17}) (blue dashed curve).  {\it Lower panel}: relative residual difference between the two curves.}
\label{Fig:4}
\end{figure}
Thus, the LC radius, $\rho_L$, satisfies the conditions $\left|F_{0,\rho}\right|=\left|F_{mc,\rho}\right|$. Taking into account (\ref{eq:7}), we obtain
\begin{equation}\label{eq:8}
\frac{\rho_{L}}{1-\rho_{L}}E(m_L)-\frac{\rho_{L}}{1+\rho_{L}}K(m_{L})=q\pi,
\end{equation}	
where $q=M_0/M_{mc}$. 
Since this equation is true in the equatorial symmetry plane $\zeta=0$, the parameter (\ref{eq:6}) of the elliptical integrals is $m_L=4\rho_L/(\rho_L+1)^2$. Equation (\ref{eq:8}) for $M_0=0$  has a trivial solution $\rho_L=0$; in this case, LC degenerates to a point coincident with the MC centre. For the case, $M_0 \gg M_{mc}$, the LC radius tends to the MC. Figure~\ref{Fig:4} ({\it solid curve}) shows the numerical solution of equation (\ref{eq:8}) for different values of the central-to-MC mass ratio, $q$. The motion around the central mass may exist in the region $\rho<\rho_L$. But in the region between LC and MC, where $\rho_L< \rho <1$, a particle shall fall toward MC (in the case of particle moving in the ring symmetry plane).

We can obtain a simpler expressions for the MC potential and, correspondingly, for the radial component of the force for the case $\zeta=0$, using the relations for the elliptical integrals \citep{Gradshteyn2007}:
\begin{equation}\label{eq:9}
K \left(
	\frac{2\sqrt{\rho}}{\rho+1}
  \right) = (1+\rho)K(\rho),
\end{equation}
\begin{equation}\label{eq:10}
E \left(
	\frac{2\sqrt{\rho}}{\rho+1}
  \right) = \frac{1}{1+\rho}
  \left[
  2E(\rho) - (1-\rho^2)K(\rho)
  \right].
\end{equation}
For simplicity, we use the module $k\equiv \sqrt{m}$ in the elliptical integrals of (\ref{eq:9}) and (\ref{eq:10}). Then, through (\ref{eq:9}), we rewrite the MC potential (\ref{eq:5}) in the equatorial plane for $\rho<1$ as:
\begin{equation}\label{eq:11}
\varphi_{mc}(\rho) = \frac{2GM_{mc}}{\pi R} K(\rho)
\end{equation} 
and, correspondingly, the radial component of gravitational force (\ref{eq:7}) for $\zeta = 0$ as:
\begin{equation}\label{eq:12}
F_{mc}(\rho)
= \frac{2GM_{mc}}{\pi R^2} \frac{1}{\rho(1-\rho^2)}
\left[
E(\rho)-(1-\rho^2)K(\rho)
\right].
\end{equation} 
Taking into account (\ref{eq:12}), the equation (\ref{eq:8}) for the LC radius simplifies in
\begin{equation}\label{eq:13}
\frac{\rho_L}{1-\rho_L^2}E(\rho_L)-\rho_LK(\rho_L)=q\frac{\pi}{2}
\end{equation}
 \citep[compare with][]{Woodward1992}. This equation can be solved analytically only for one value of the mass ratio with the help of the relations \citep{Gradshteyn2007}
\[
K\left(\frac{\sqrt{2}}{2}\right)=\frac{\Gamma(1/4)^2}{4\sqrt{\pi}}; \quad
E\left(\frac{\sqrt{2}}{2}\right)=\frac{\pi^{3/2}}{\Gamma(1/4)^2}+
\frac{\Gamma(1/4)^2}{8\sqrt{\pi}}.
\]  
Then, for $\rho_L=\sqrt{2}/2$, the solution of equation (\ref{eq:13}) gives the following value of the mass ratio $q\equiv M_0/M_{mc}=2\sqrt{2\pi}\Gamma(1/4)^{-2}\approx 0.38$.

\subsection{Approximate solutions for the MC potential and radius of Lagrangian circle}
\label{subsec:Approximate}

We will construct here an approximate expression of the MC potential for $\rho<1$ which allows us to obtain an analytical solution for the LC radius. For $\rho \rightarrow 0$, $K(0) \rightarrow \pi/2$ and 
\[
\left.\varphi_{mc}\right|_{\rho\rightarrow 0} \approx \frac{G M_{mc}}{R}.
\] 
For   
$\rho \rightarrow 1$, $m \rightarrow 1$, $K(m) \rightarrow \ln (4/\sqrt{1-m}) = \ln(4(1+\rho)/(1-\rho)) \approx \ln(8/(1-\rho))$, and
\[
\left.\varphi_{mc}\right|_{\rho\rightarrow 1}\approx \frac{G M_{mc}}{\pi R}\ln\frac{8}{1-\rho}.
\]
Sewing these limiting cases, we obtain a good approximation for the MC potential for $\rho <1$:
\begin{equation}\label{eq:14}
\varphi_{mc}(\rho) \approx \frac{GM_{mc}}{R}\left[1-\frac{1}{\pi}\left(\rho + \ln(1-\rho)\right)\right],
\end{equation}  
and, for the radial component of the gravitational force: 
\begin{equation}\label{eq:15}
F_{mc,\rho}(\rho) \approx \frac{GM_{mc}}{\pi R^2}\frac{\rho}{1-\rho},
\end{equation} 
which is simpler than the expression (\ref{eq:12}).
\begin{figure}
\includegraphics[width = 80mm]{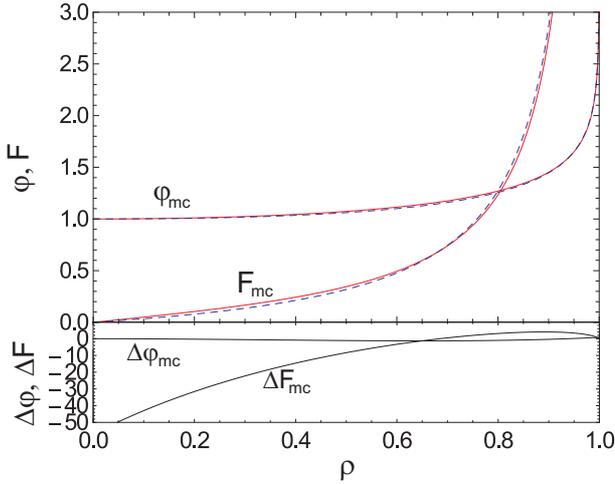}
\caption{{\it Upper panel}: dependence of the potential and the radial component of the gravitational force for a given MC: the solid curves correspond to the exact expressions (\ref{eq:11}), (\ref{eq:12}), and the dashed curves to the approximated case (\ref{eq:14}), (\ref{eq:15}) for $M_0=M_{mc}$. {\it Lower panel}: relative residual differences between two such curves.}
\label{Fig:5}
\end{figure} 

The curves of the MC potential and of the radial component of the force, calculated by the exact expressions (\ref{eq:11}) and (\ref{eq:12}) and though of the approximated expressions (\ref{eq:14}) and (\ref{eq:15}), are shown in Fig.~\ref{Fig:5}. The approximated expression for the potential works with a good precision in the whole region under consideration ($\rho < 1$).
This is no longer true for their gradient.  Nonetheless there is a good correspondence between the exact and approximated expressions for the force in the region $\rho \gg 0$ which is of interest for the following considerations.
By equating the gravitational forces from the central mass and from the MC using the approximated expression (\ref{eq:15}), we obtain the equation for the LC:
\begin{equation}\label{eq:16}
\rho_L^3 - q\pi(1-\rho_L)= 0.
\end{equation}  
This equation provides the analytical solution for the LC radius
\begin{equation}\label{eq:17}
\rho_L = \frac{\pi^{1/3}(2^{1/3} s^2 -2q (3\pi)^{1/3})}{6^{2/3}s},
\end{equation}  
where
\[
s=(9q + \sqrt{3}q\sqrt{27 + 4 \pi q})^{1/3}.
\]
It is apparent from Fig.~\ref{Fig:4} that there is a good agreement between the numerical solution of the exact LC equation (\ref{eq:13}) and the approximated one (\ref{eq:17}).

\section{THE MOTION IN THE MC EQUATORIAL PLANE}
\label{sec:OSCO}

Let us consider the problem of the existence of finite and, in particular, circular orbits for a particle moving in the equatorial plane under the effect of the gravitational field of  a MC with a central mass. This problem is the particular case of the particle motion in an axisymmetric potential 
\citep[and references therein]{Binney1987, Ninkovic2009}
and differs from the previous ones because the gravitational forces by the central mass and by the MC act in opposite directions. The effective potential is
\begin{equation}\label{eq:18}
U_{eff}=U_0+U_{mc}+\frac{I^2}{2 R^2\rho^2},
\end{equation}
where $U_0=-GM_0/(R\rho)$ and $U_{mc}=-\varphi_{mc}$ are the potential energies of the central mass and of the MC, and $I=I_\zeta$ is the angular momentum per unit particle mass.
\begin{figure}
\includegraphics[width = 80mm]{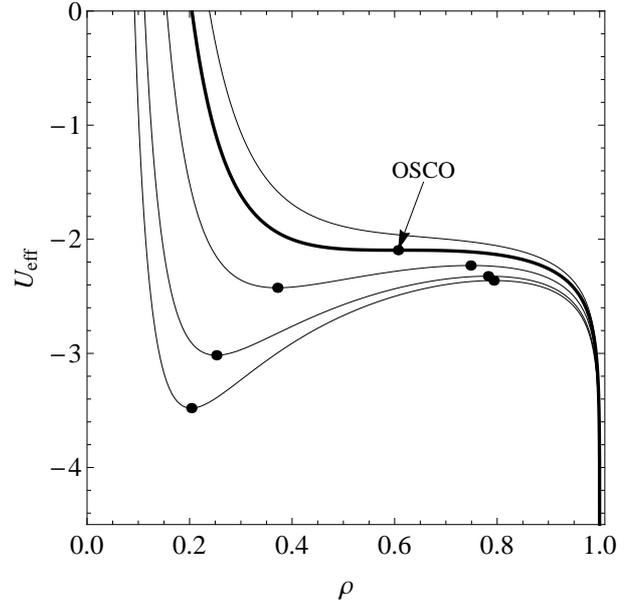}
\caption{Radial dependence of the effective potential for different values of the angular momentum: $I=0.45$, $0.5$, $0.6$, $0.702$, $0.77$ and $M_0=M_{mc}$.  Extrema of $U_{eff}$ are shown by solid dots; the thick curve corresponds to $I=I_{osco}=0.702$.} 
\label{Fig:6}
\end{figure}
\begin{figure}
\includegraphics[width = 80mm]{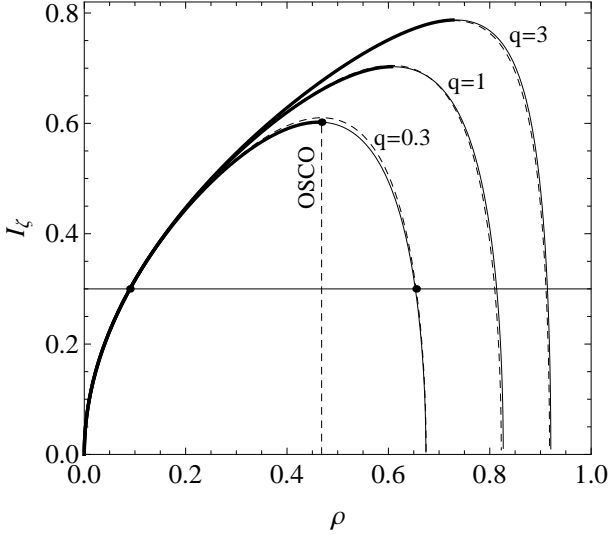}
\caption {Graphical solution of the equation (\ref{eq:19}) taking into account the exact expression (\ref{eq:20}) (solid curves) and the approximation  (\ref{eq:21}) (dashed curves) for $I=0.3$ and the different core-to-MC mass ratios $q=0.3, 1, 3$.  The thick curves correspond to the region where the stable circular orbits exist.}
\label{Fig:7}
\end{figure}

For small values of the angular momentum, $U_{eff}$ has two extrema: a minimum, which 
corresponds to a stable circular orbit, and a maximum, which corresponds to an unstable 
circular orbit (Fig.~\ref{Fig:6}). By increasing the value of the momentum, the minimum shifts 
towards higher $\rho$ and the maximum toward lower $\rho$. For some boundary value of the 
angular momentum, the minimum and the maximum merge with each other into one point. This point 
corresponds to the maximal possible radius of the stable circular orbit. In general, such behavior of $U_{eff}$ is similar to the case of particle motion around a black hole \citep{Landaufshitz1975, Abramowicz2013, Tsupko2016}. Indeed, in the relativistic case the effective potential has also two extrema: maximum at lower radii and minimum at higher ones. By increasing the angular momentum, the extrema of the effective potential merge into one point which also corresponds to the last possible stable circular orbit \citep[see][figure~2]{Tsupko2016}. 
This orbit is called the innermost stable circular orbit (ISCO) because the stable circular orbits are located outside of the ISCO. For Schwarzschild metric, the radius of the ISCO equals $3R_g$, where $R_g=2GM/c^2$ is the Schwarzschild radius. In the potential considered here the situation is opposite to the relativistic case: the stable circular orbits are located inside of the last stable circular orbit. By analogy with relativistic case, we will call this orbit "the outermost stable circular orbit" (OSCO).  

\begin{figure}
\includegraphics[width = 80mm]{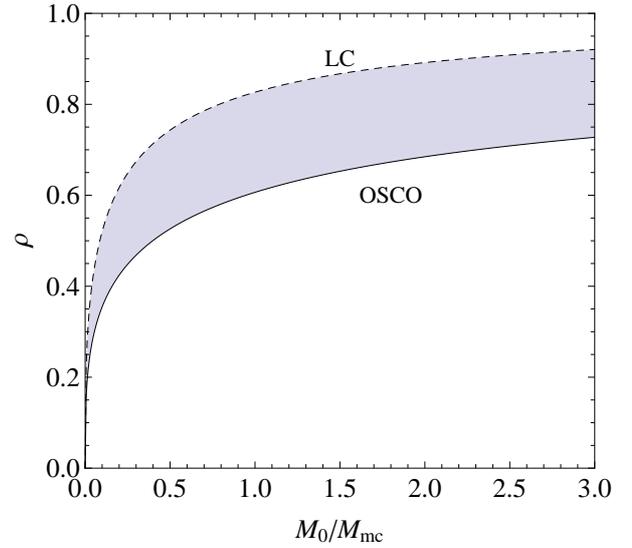}
\caption {Dependence of the OSCO and of the LC radii on core-to-MC mass ratio. The gray color marks the region where the stable circle orbits do not exist.}
\label{Fig:8}
\end{figure}

The extrema of the effective potential, $\partial U_{eff}/\partial \rho$, lead us to the following equation:
\begin{equation}\label{eq:19}
I^2=W(\rho),
\end{equation}
where
\[
W(\rho) = \rho+\frac{1}{q \pi}\frac{\rho^2}{\rho^2-1}
\left[
(\rho+1)E(m)+(\rho-1)K(m)
\right]=
\]
\begin{equation}\label{eq:20}
\hspace*{0.8cm} = \rho+\frac{1}{q \pi}2\rho^2
\left[
K(\rho)-\frac{E(\rho)}{1-\rho^2}
\right].
\end{equation}
To solve equation (\ref{eq:19}) we use the approximated expression (\ref{eq:14}) for MC holding in the inner region $\rho<1$:
\begin{equation}\label{eq:21}
W(\rho) \approx \rho + \frac{\rho^3}{q\pi} \left(1-\frac{1}
{1-\rho}\right). 
\end{equation}
A graphic solution of equation (\ref{eq:19}) in Fig.~\ref{Fig:7} ({\it thick curves}) shows  the range of radii for which circular orbits exist. 
For the same value  $I < I_{osco}$, there are two solutions for $\rho$ which correspond to existence of the stable circular orbits (smaller $\rho$) and of unstable circular orbits (larger $\rho$). So, the maximal possible value of the momentum, $ I_{osco}$, discriminates two region of existence of the stable circular orbits and of unstable equilibrium points. The equation for the OSCO can be obtained from the condition $\partial W(\rho)/\partial \rho =0$ by taking into account (\ref{eq:21}):
\begin{equation}\label{eq:22}
3\rho_{osco}^4 - 4 \rho_{osco}^3 + \pi q (1-\rho_{osco})^2=0.
\end{equation}
This equation has two real roots, one of which satisfies the condition $\rho < 1$. For example,  the radius of the OSCO for $q=1$ is  $\rho_{osco} \approx 0.606$ from the numerical solution of equations (\ref{eq:22}), and the angular momentum of the OSCO $I_{osco} \approx 0.703$ from (\ref{eq:19}). The dependence of the OSCO and of the LC radii on the core-to-MC mass ratio are shown together in Fig.~\ref{Fig:8}.
We can test the disappearing of stable circular orbits by solving the equation of motion: 
\begin{equation}\label{eq:23}
\ddot{\vec{\rho}} =-\frac{G M_{0}}{R^3} \vec{\rho}
	\left[
    \frac{1}{(\rho^2+\zeta^2)^{3/2}} -\frac{f(\rho,\zeta)}{\pi q \, \rho^{5/2}}  
    \right] 
\end{equation}
\begin{equation}\label{eq:24}
\ddot{\zeta} =- \frac{G M_0}{R^3}\zeta
		\left[		
		\frac{1}{(\rho^2+\zeta^2)^{3/2}} +			
					\frac{1}{\pi q }\sqrt{\frac{m}{\rho}}\frac{E(m)}{(\rho -1)^2+\zeta^2}    \right] 
\end{equation}
where
\[
f(\rho,\zeta)=\frac{\sqrt{m}}{4(1-m)}
\left[
(2-m(\rho+1))E(m)-2(1-m)K(m)
\right],
\]
$\zeta$, $\vec{\rho}=(x,y)$  are parametrized by the ring radius $R$; the parameter $m$ of the elliptical integrals is defined by the expression (\ref{eq:6}). In the polar coordinates ($\rho$, $\psi$), the equations of motion in the equatorial plane are:
\begin{equation}\label{eq:25}
\ddot{\rho} - \rho \dot{\psi}^2=-\frac{G M_{0}}{R^3 \rho^2} 
	\left[
    1 -\frac{1}{\pi q}\sqrt{\rho} f(\rho,0)  
    \right] 
\end{equation}
\begin{equation}\label{eq:26}
\dot{\psi}\rho^2 R^2 = I
\end{equation}
\begin{figure*}\centering
\includegraphics[width = 55mm]{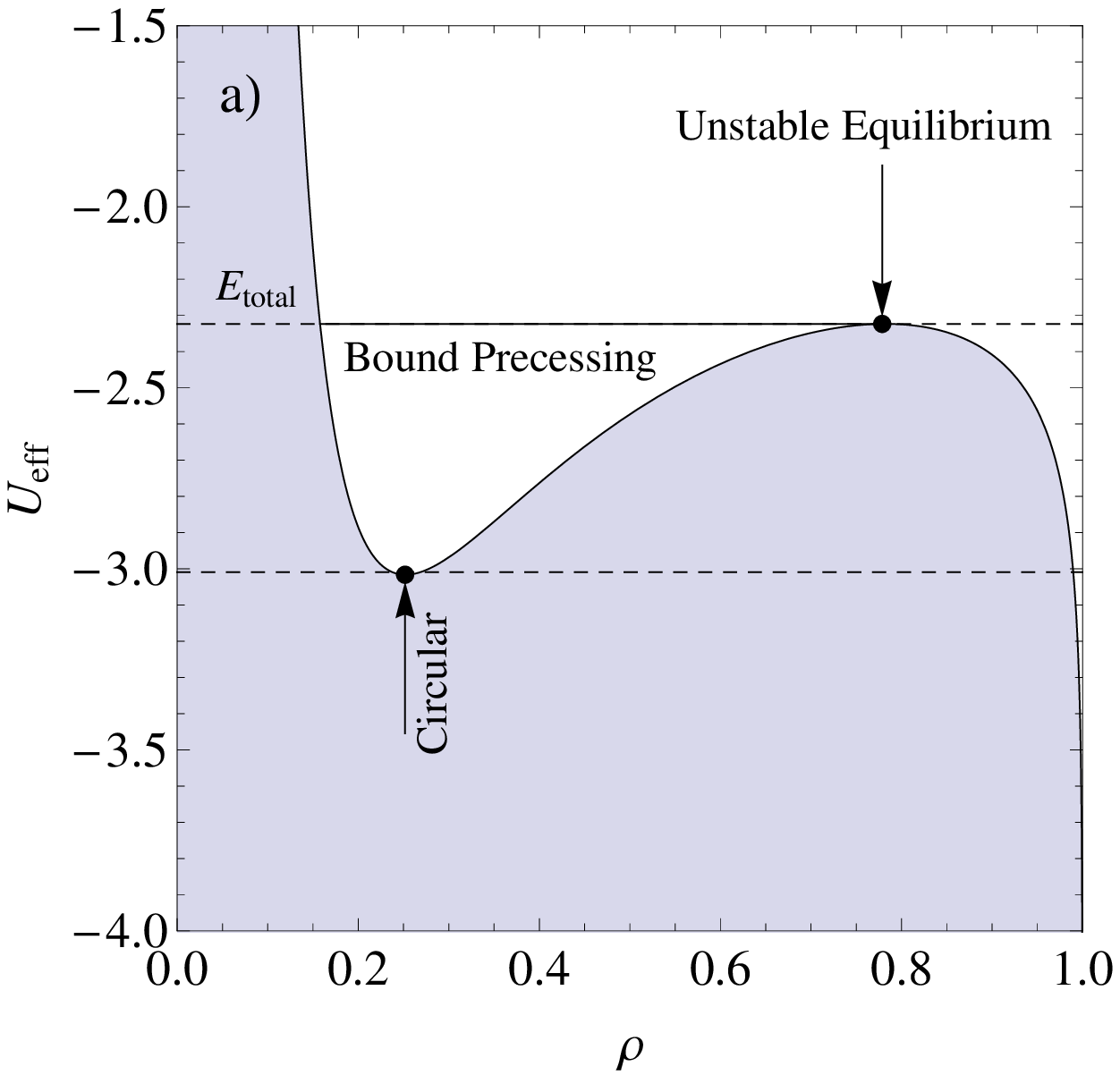}
\includegraphics[width = 55mm]{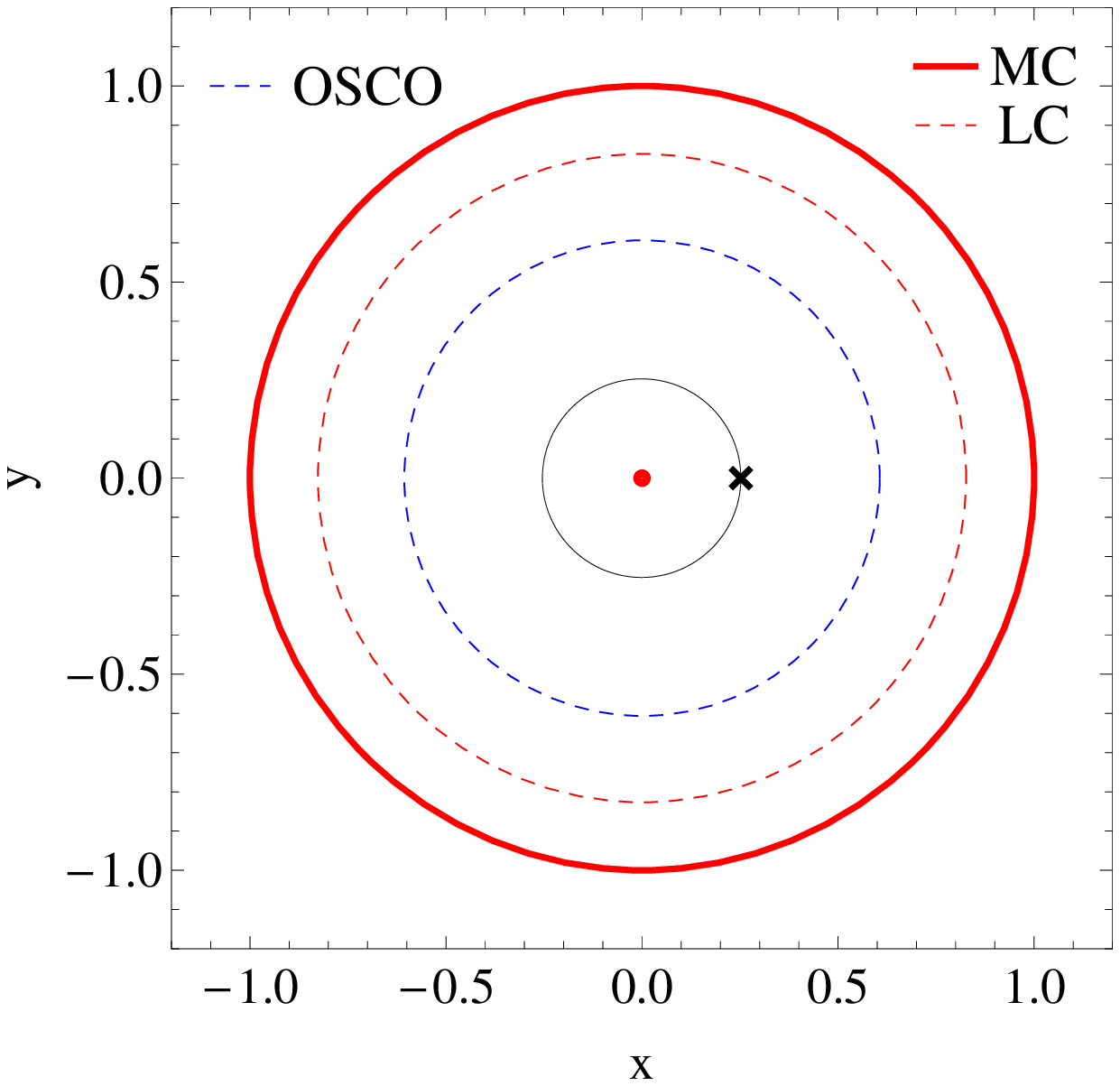}
\includegraphics[width = 55mm]{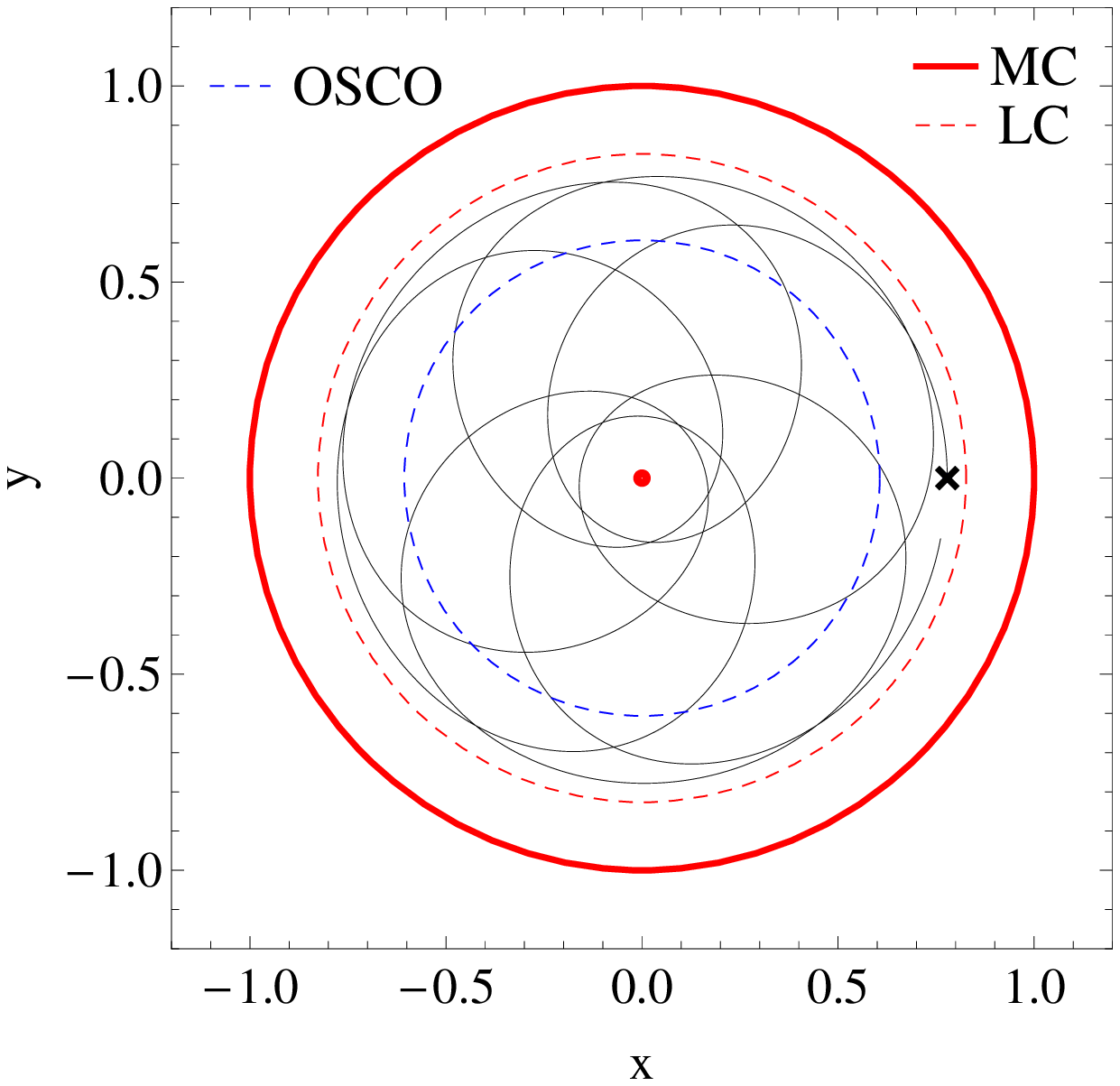}
\includegraphics[width = 55mm]{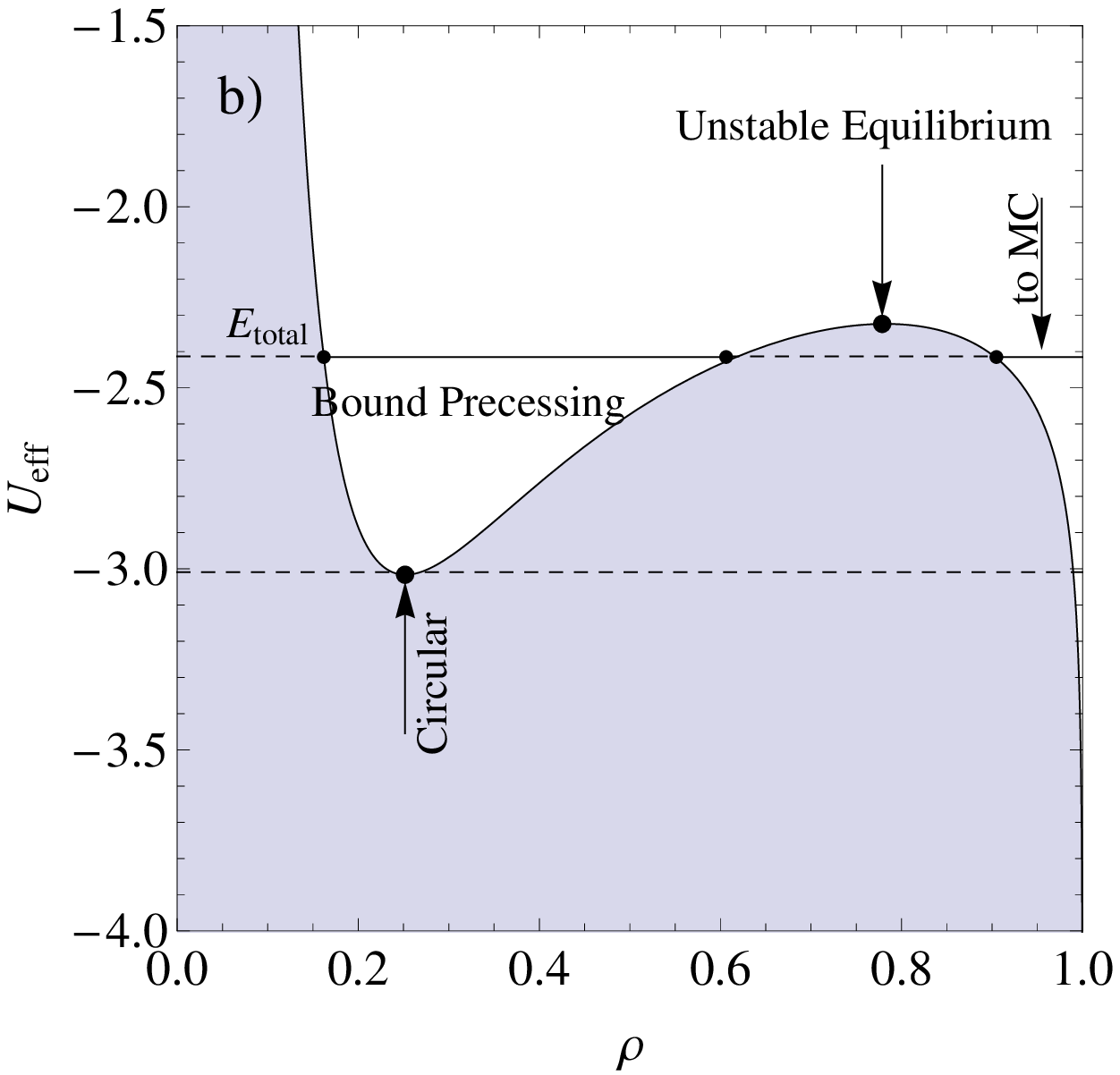}
\includegraphics[width = 55mm]{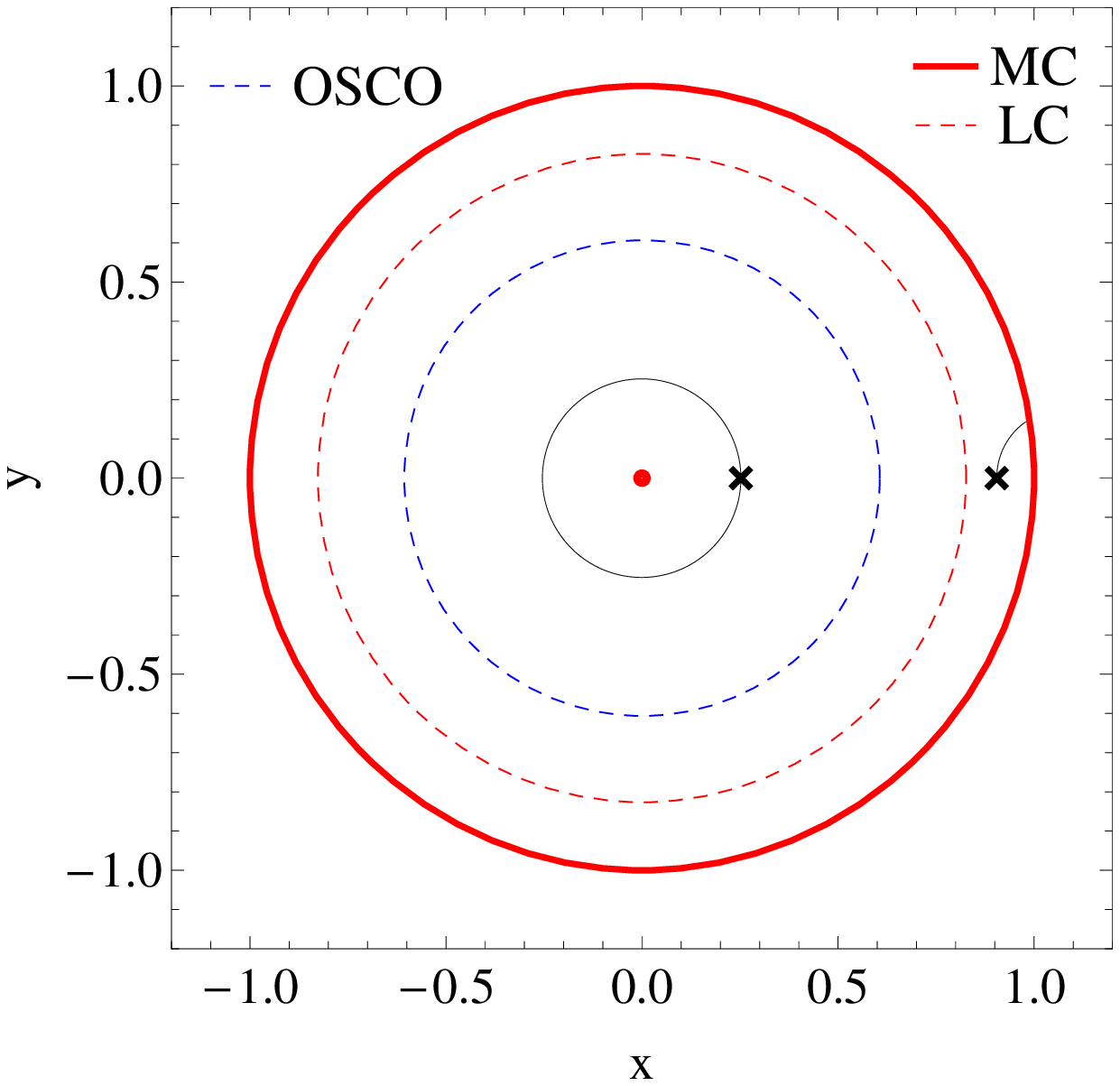}
\includegraphics[width = 55mm]{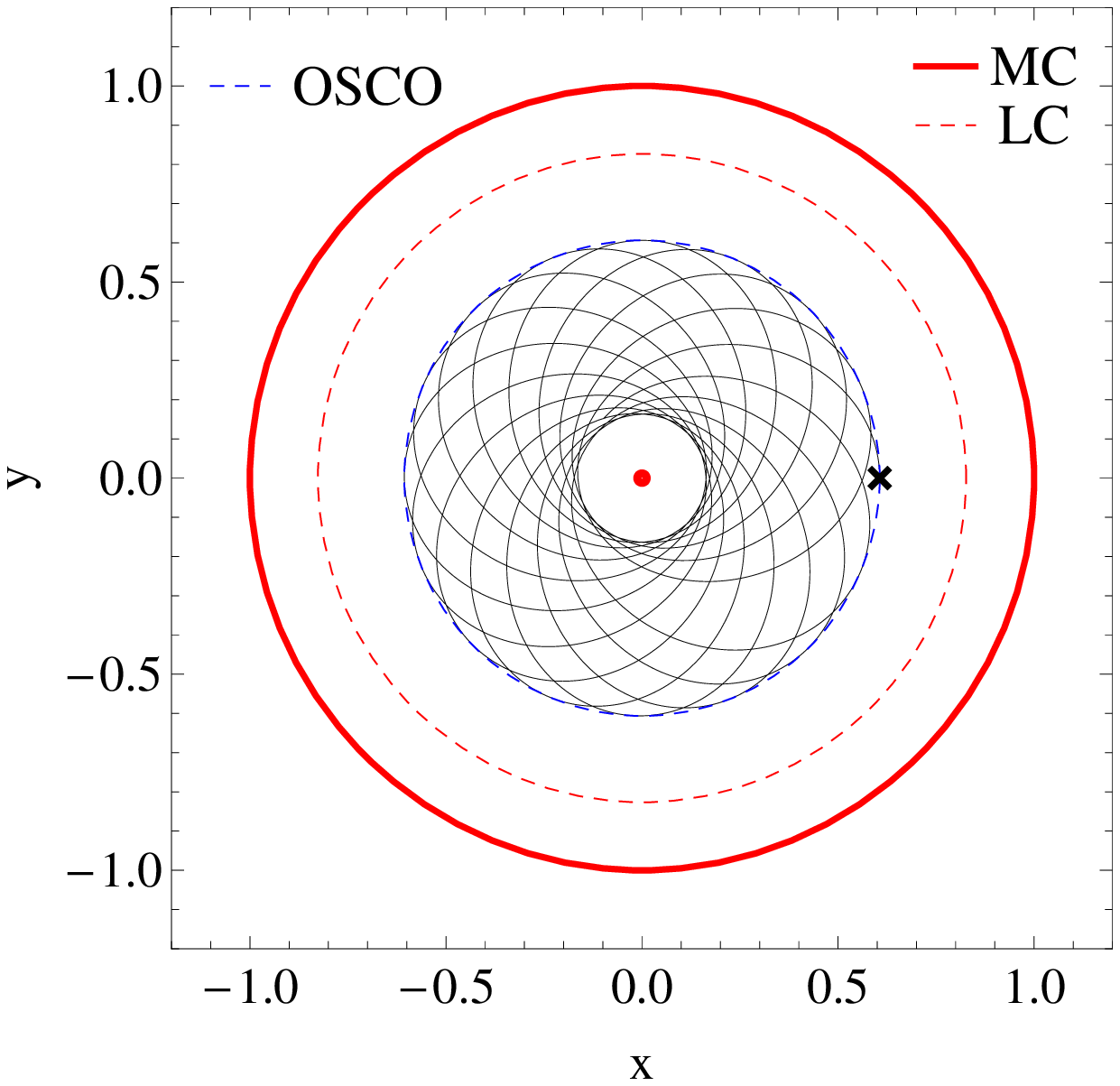}
\includegraphics[width = 55mm]{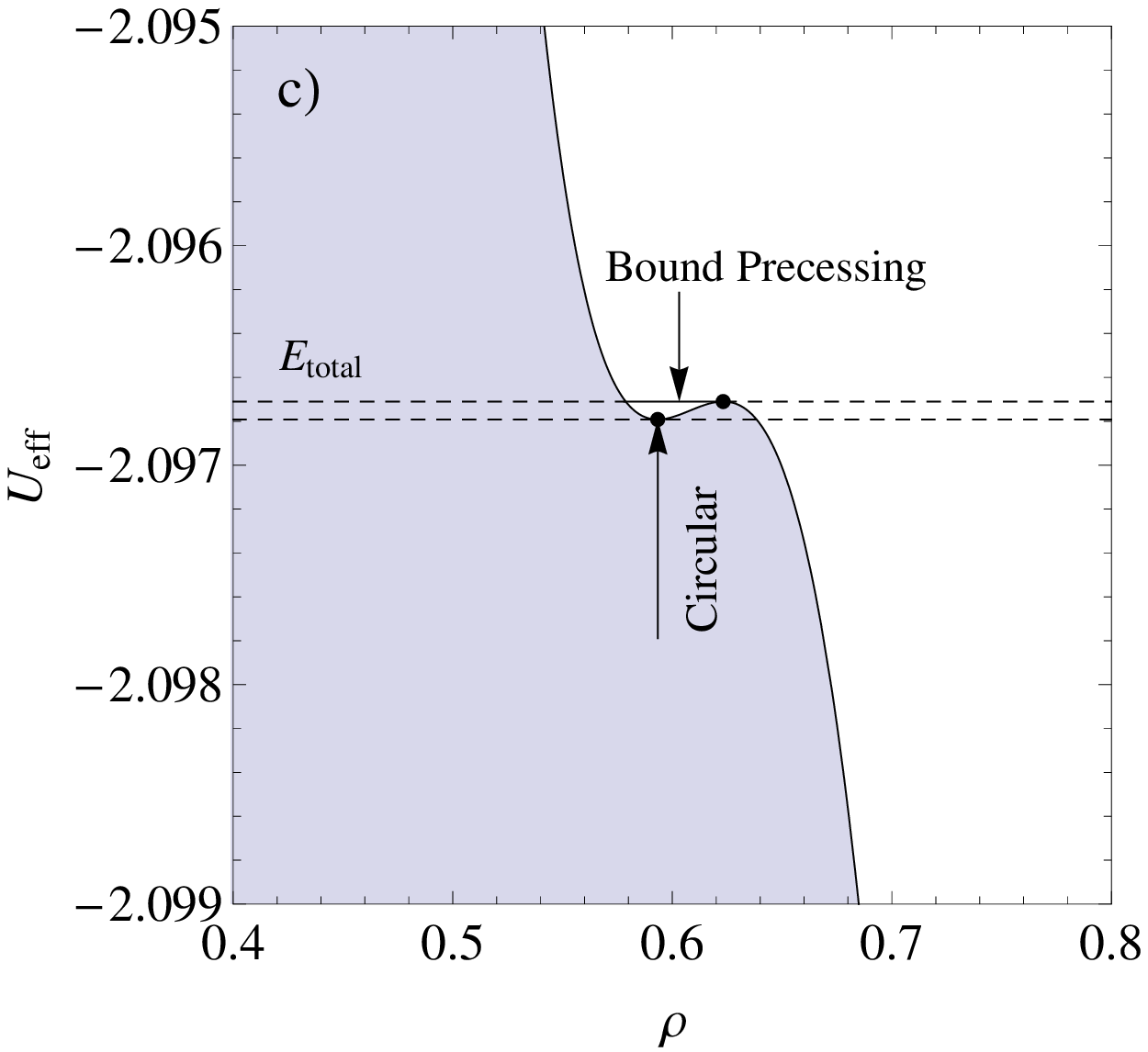}
\includegraphics[width = 55mm]{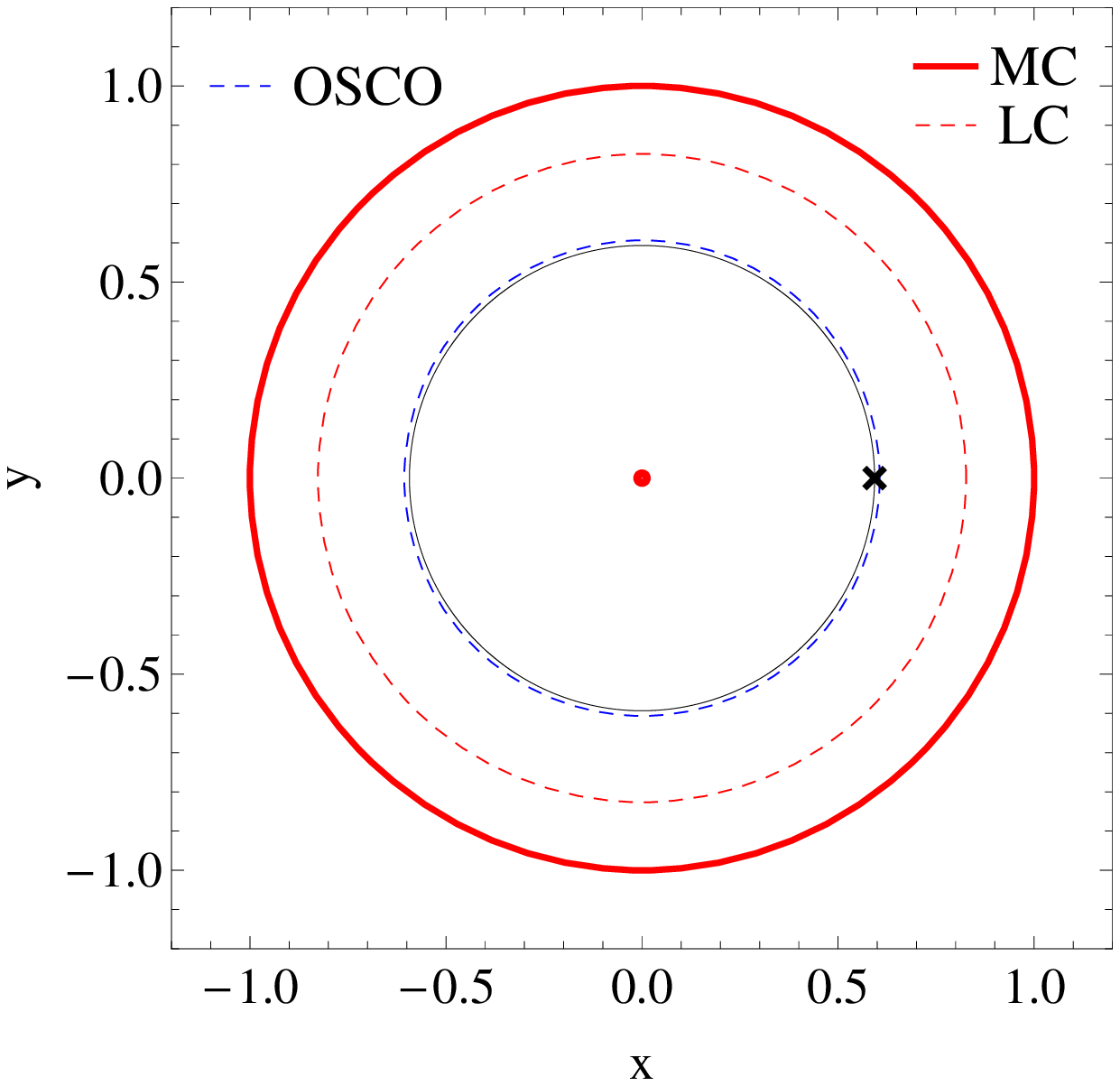}
\includegraphics[width = 55mm]{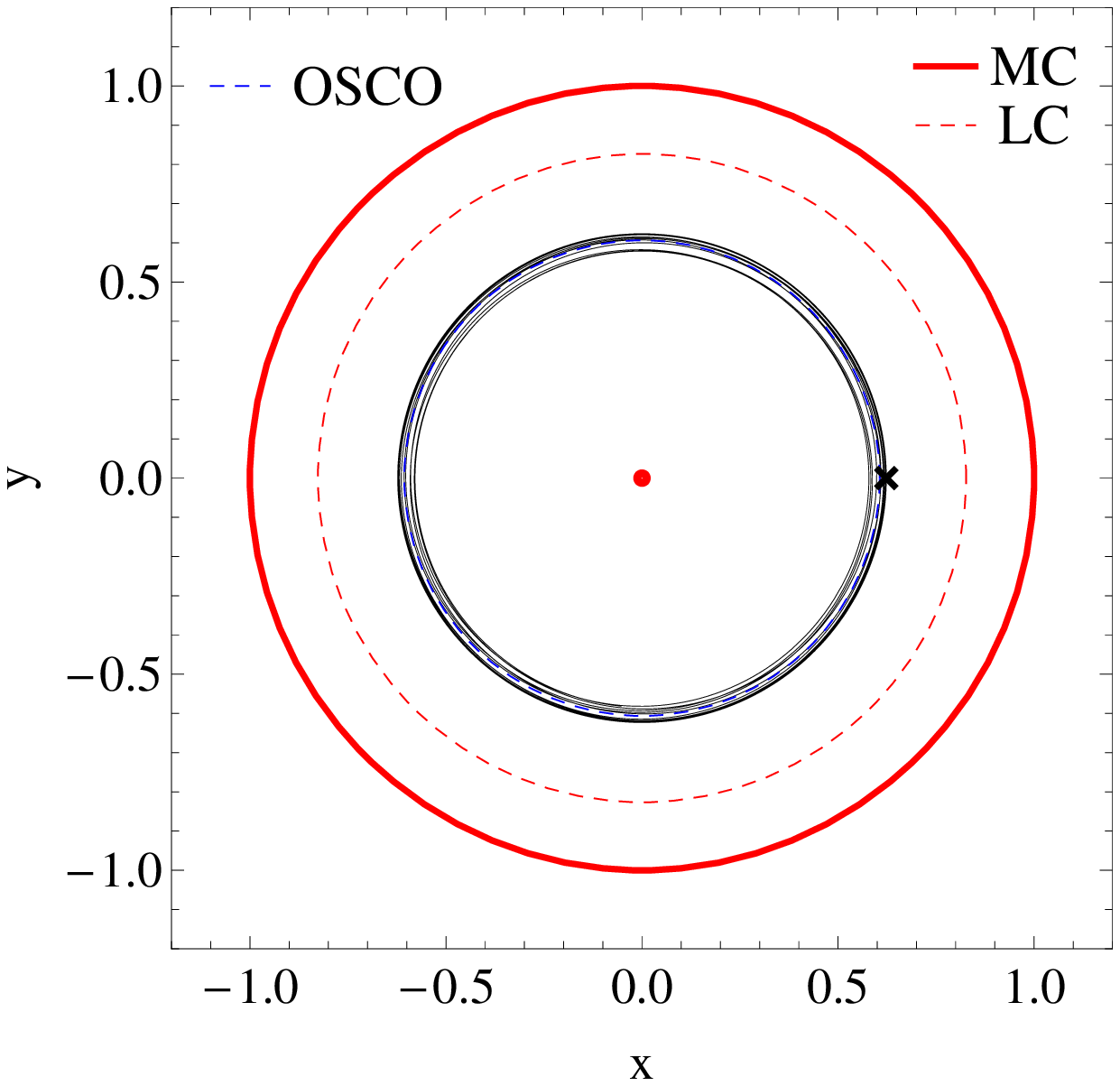}
\includegraphics[width = 55mm]{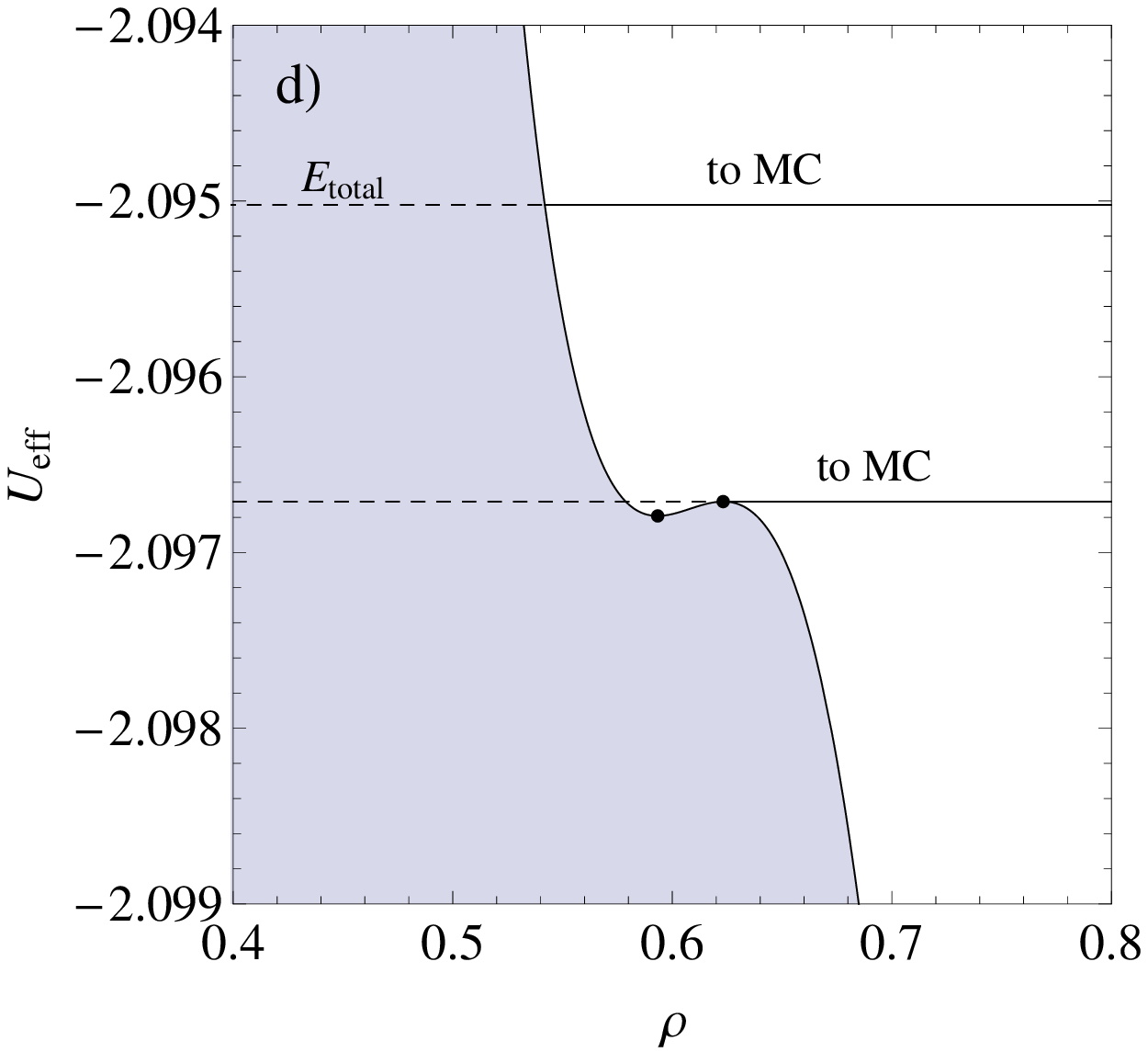}
\includegraphics[width = 55mm]{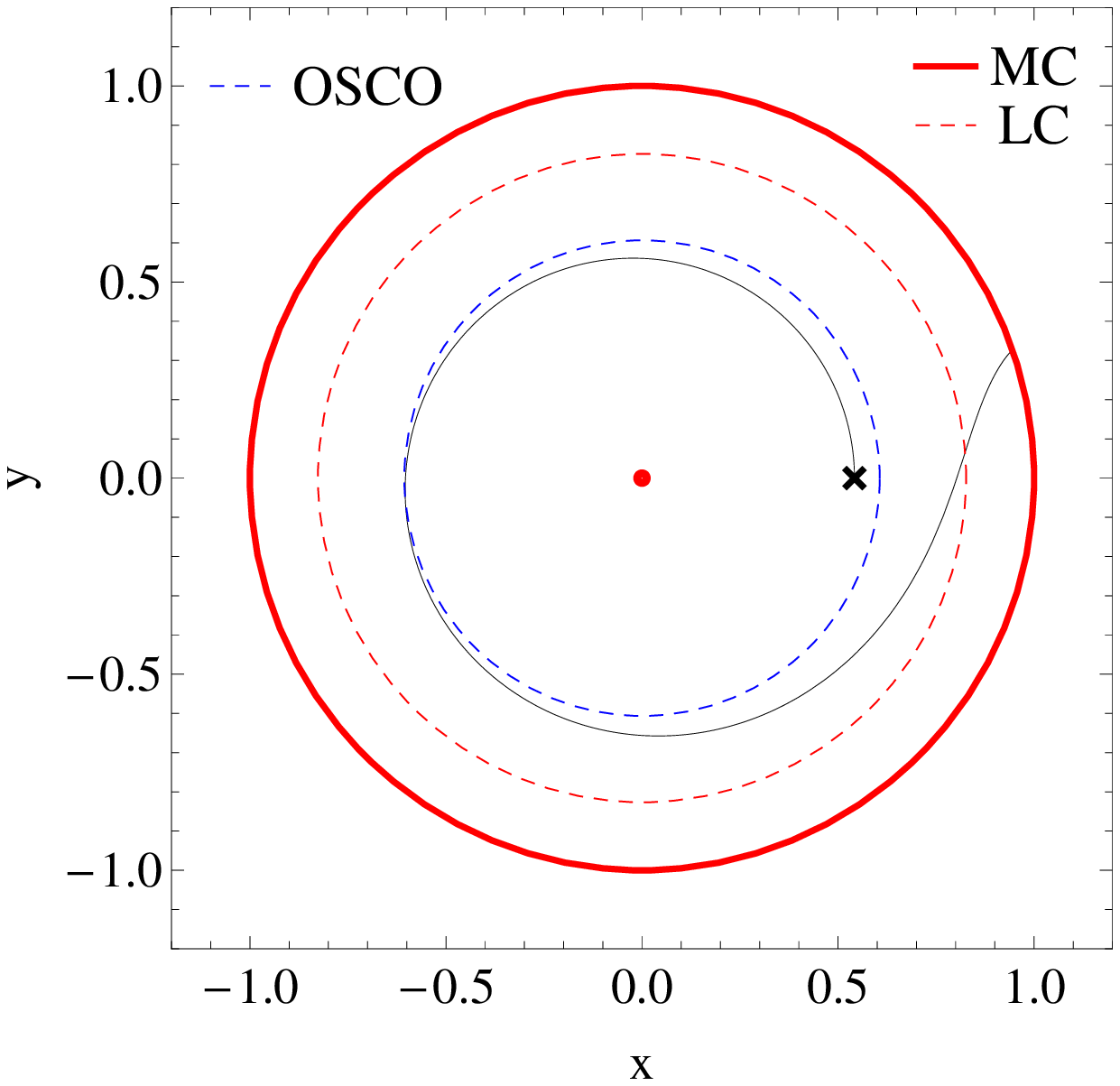}
\includegraphics[width = 55mm]{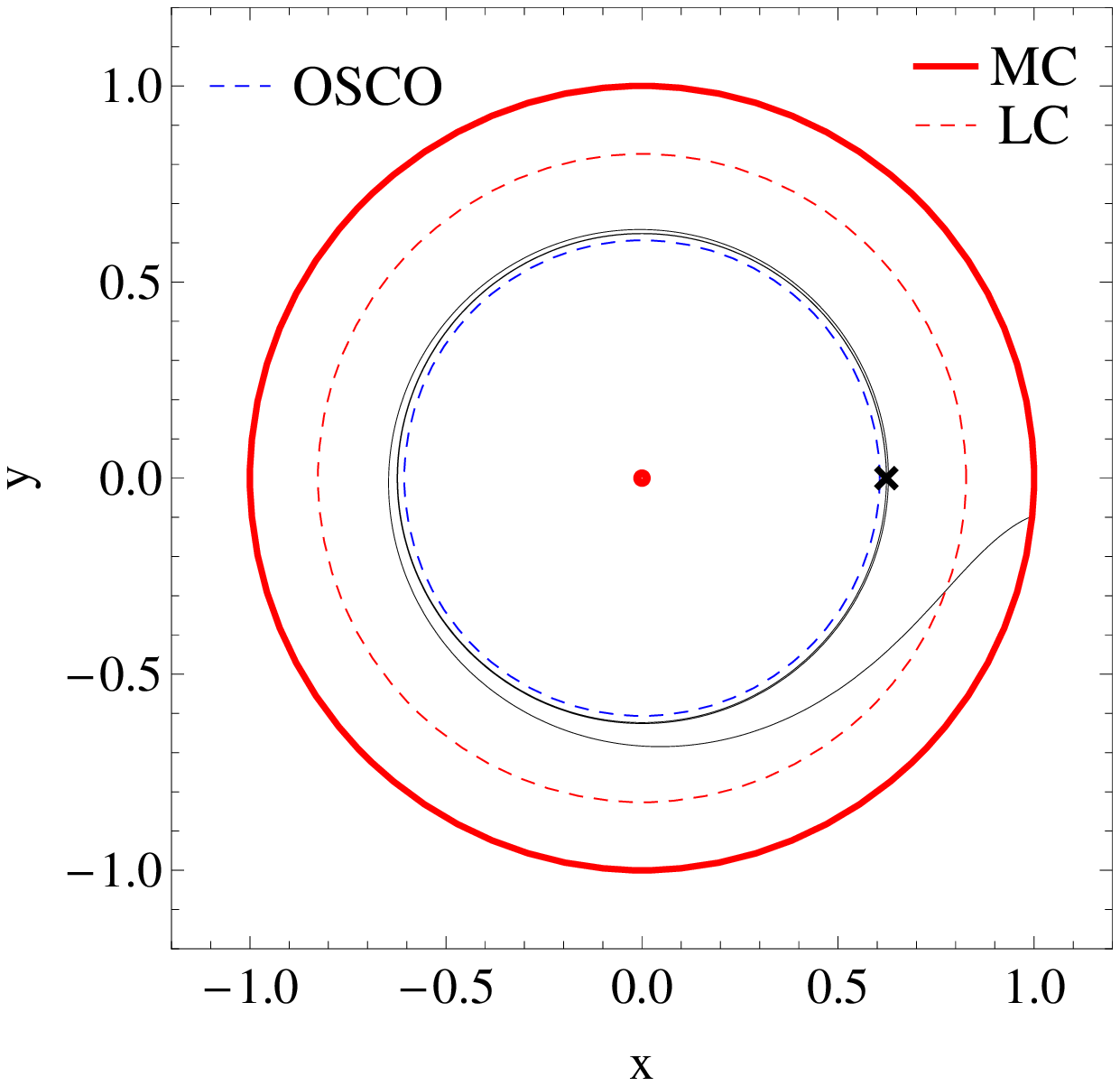}
\caption {The effective potential ({\it left columns}) and orbits of a test particle  ({\it medium, right columns}) in the MC equatorial plane for $M_0=M_{mc}$. The initial conditions: a) $I=0.5$, $\rho_0=0.252$, $\rho_0=0.779$; b) $I=0.5$, $\rho_0=0.252$, $\rho_0=0.606$; c) $I=0.702$, $\rho_0=0.593$, $\rho_0=0.623157$; d) $I=0.702$, $V_\rho(0)=1.295$, $\rho_0=0.623220$. 
The cross marks the starting point of the orbit. Solid red, dashed red, and dashed blue circles represent the MC, the LC, and the OSCO, correspondingly.
}
\label{Fig:9}
\end{figure*}
Examples of the orbits with the different initial conditions are shown in Fig.~\ref{Fig:9}. For $I<I_{osco}$, the effective potential has a minimum, $\min(U_{eff})$, which corresponds the stable circular orbit (Fig.~\ref{Fig:9}~a,b,c; {\it medium columns}). The unstable equilibrium is realized for $E_{total}=\max(U_{eff})$ and in this case a particle moves on a bound precessing (rosette) orbit (Fig.~\ref{Fig:9}~a; {\it right columns}). The extrema of $U_{eff}$ were found by solving equation (\ref{eq:19}). If $E_{total}<\max(U_{eff})$, there are bound precessing orbits in the potential well and also motions towards MC (Fig.~\ref{Fig:9}~b; {\it medium, right columns}). For $I \rightarrow I_{osco}$, the extremes of $U_{eff}$ vanish, and the radius of the stable circular orbit tends to $\rho_{osco}$ (Fig.~\ref{Fig:9}~c; {\it medium column}). In this case, for $E_{total}= \max(U_{eff})$ the orbit precesses near $\rho_{osco}$ (Fig.~\ref{Fig:9}~c; {\it right column}). Since the extrema near to each other, a little deviation from previous initial condition, $E_{total} >\max(U_{eff})$, leads to an increase of the orbit radius and to motion of a particle toward the MC (Fig.~\ref{Fig:9}~d; {\it right column}). If $I>I_{osco}$, the orbits around the central mass do not exist, and particle always moves towards the MC as in Fig.~\ref{Fig:9}~d~({\it medium column}).

The region $\rho_{osco} < \rho <\rho_{L}$, where there are not stable circular orbits, is a result of the competitions between  the gravitational forces of the central mass and of the MC.  Stable circular motion can be established only within the limits of the OSCO. In the object with a massive ring, the stellar matter could be thrown out from the region with $ \rho >\rho_{osco}$. These processes might lead to a sweeping of this region and to the formation of a gap in the matter distribution.  Such a gap might explain the minimum in the star distribution between the central object and the stellar ring in such ring galaxies as Hoag's Object (see Sec.~\ref{sec:discussion}). 

\section{MOTION OF A PARTICLE IN THE MC MERIDIONAL PLANE}
\label{sec:merid}

\begin{figure}
\includegraphics[width = 79mm]{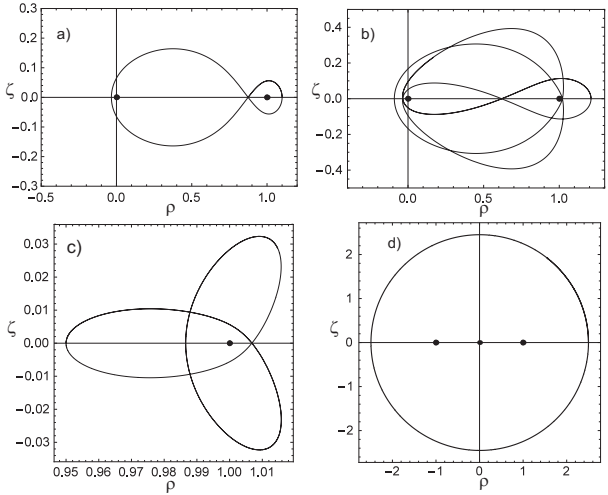}
\caption {Closed trajectories of a test particle in the MC meridional plane for $M_0=M_{mc}$ and the initial conditions: $\zeta_0=0$, $V_\rho(0)=0$, a) $\rho_0=1.1$, $V_\zeta(0)=0.392$, b)$\rho_0=1.21$, $V_\zeta(0)=0.3924$, c) $\rho_0=0.95$, $V_\zeta(0)=0.145$, d) $\rho_0=2.5$, $V_\zeta(0)=0.905$. The central mass is located in the point (0,0) and the coordinates of the MC  are ($\pm 1, 0$).}
\label{Fig:10}
\end{figure}
\begin{figure}
\includegraphics[width = 85mm]{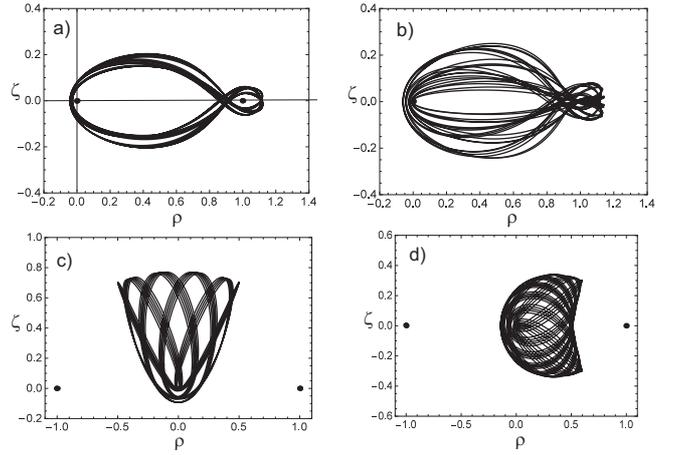}
\caption {Regular orbits of a test particle in gravitational field of the central mass and MC for  $M_0=M_{mc}$ and the initial conditions: $\zeta_0=0$, $V_\rho(0)=0$, a) $\rho_0=1.11$, $V_\zeta(0)=0.392$, b) $\rho_0=1.12$, $V_\zeta(0)=0.392$, c) $\rho_0=0.5$, $\zeta_0=0.7$, $V_\zeta(0)=0$, d) $\rho_0=0.6$, $\zeta_0=0.3$,$V_\zeta(0)=0$. The central mass is located in the point (0,0) and the coordinates of the MC are ($\pm 1, 0$).}
\label{Fig:11}
\end{figure}
\begin{figure*}\centering
\includegraphics[width = 80mm]{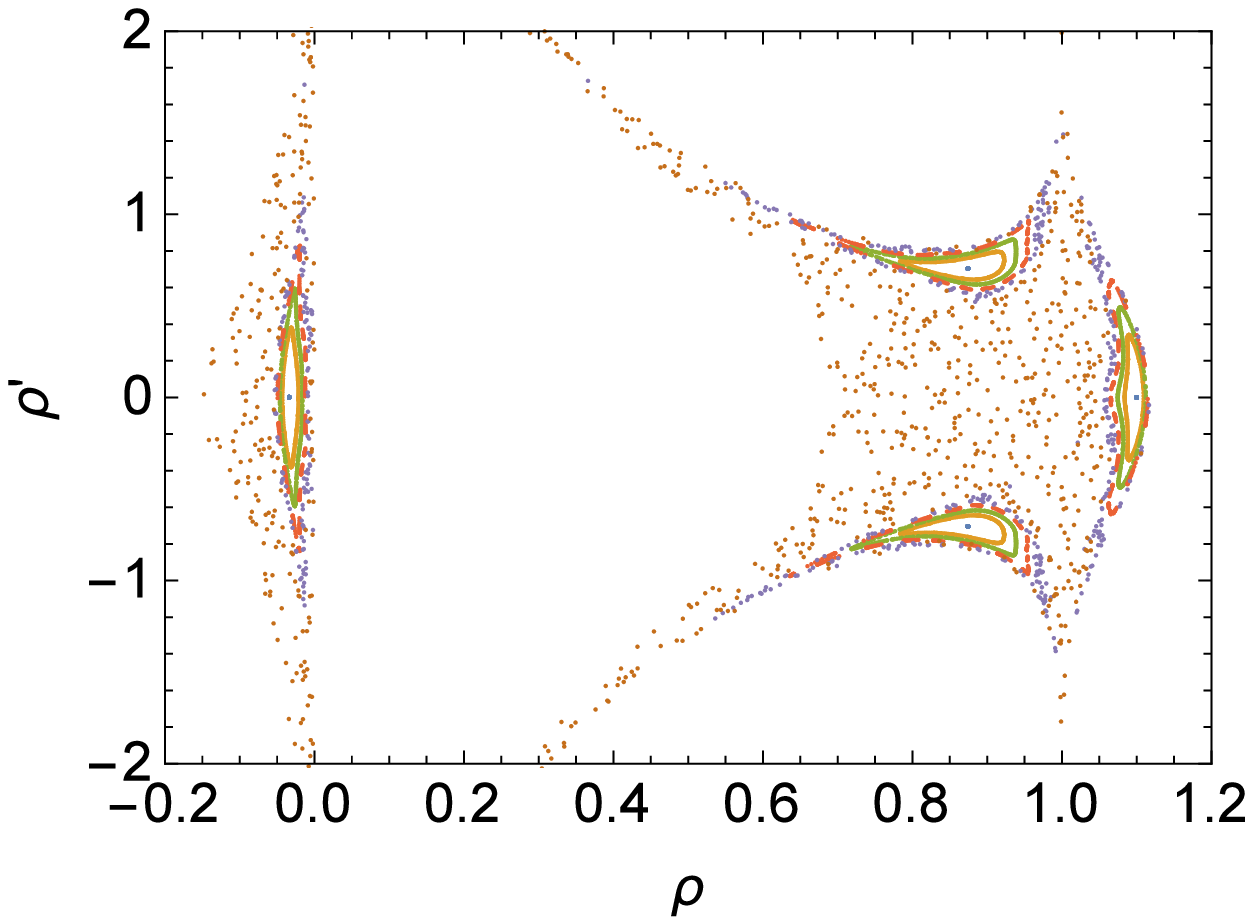}
\includegraphics[width = 80mm]{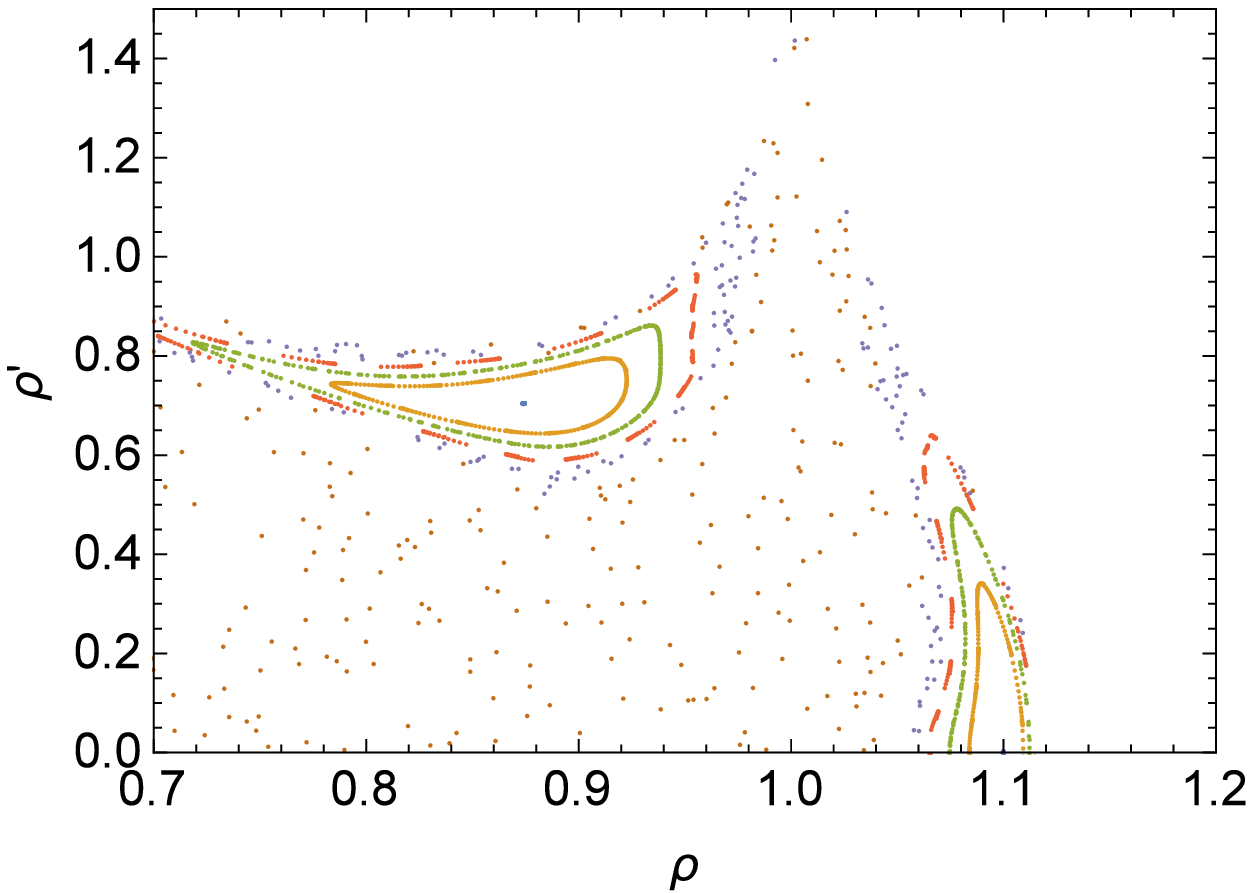}
\includegraphics[width = 80mm]{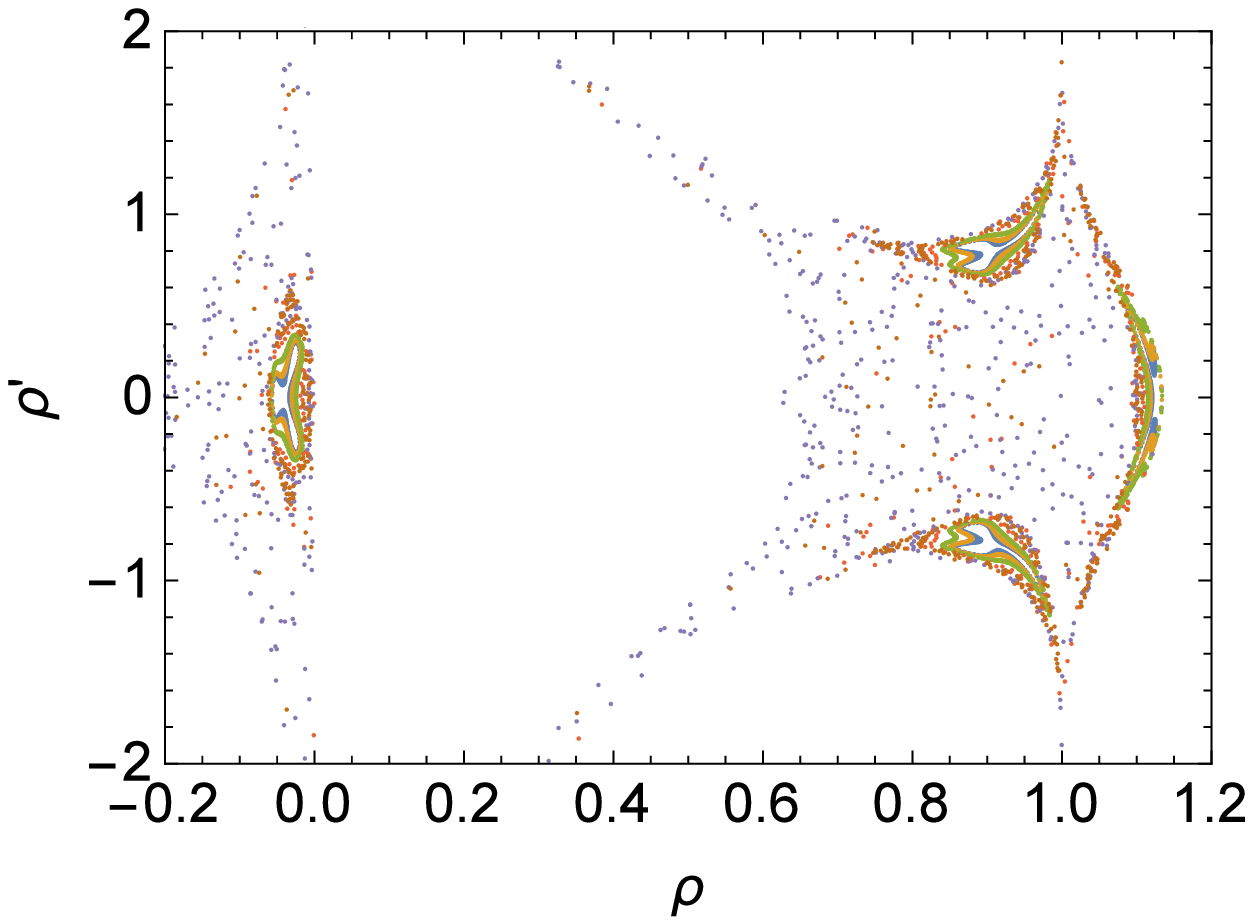}
\includegraphics[width = 80mm]{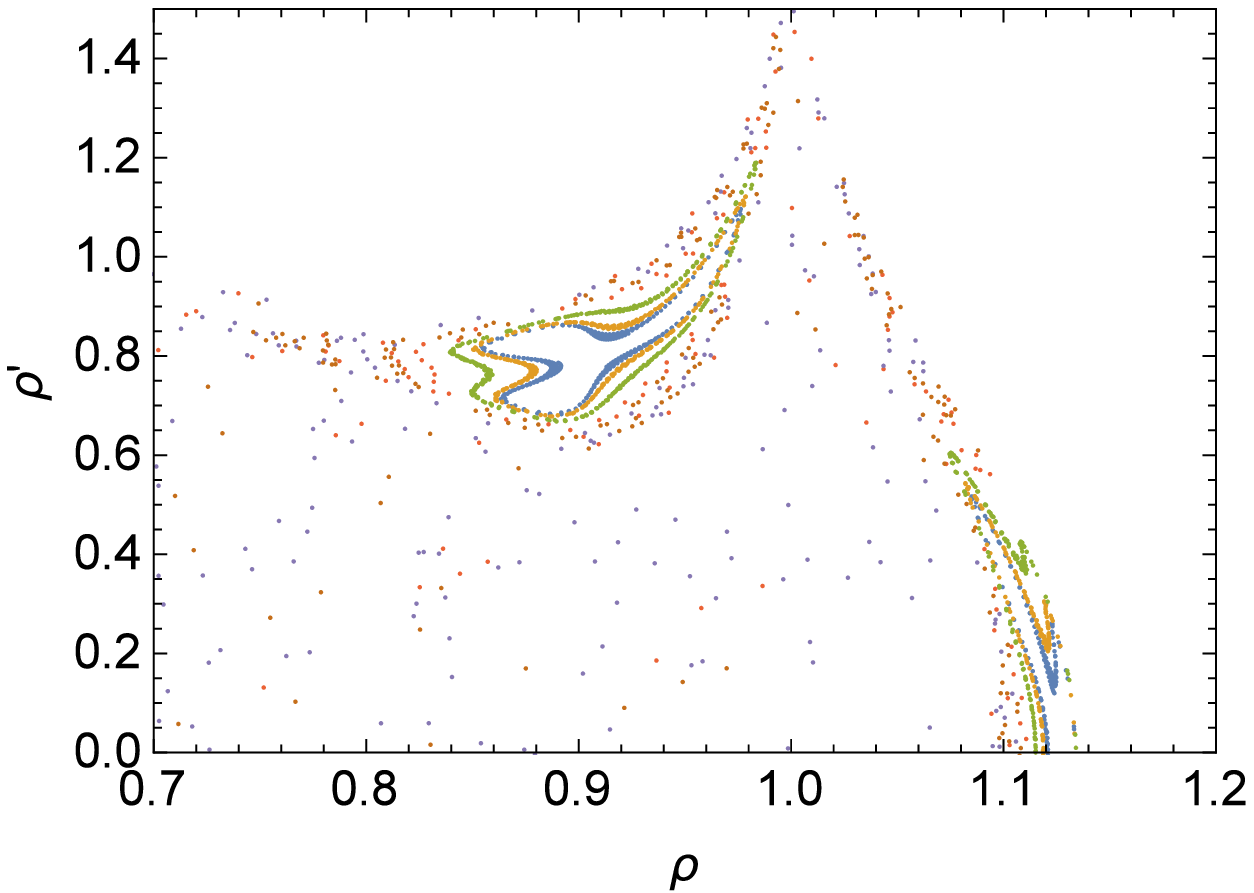}
\caption {Poincar\'e section for the initial conditions: $M_0=M_{mc}$, $\zeta_0=0$, $V_\zeta(0)=0.392$, $\rho_0=1.1 - i \cdot 0.012$, $i=0,..,5$. {\it Top panels}: $E_{total}=-2.17606$; {\it bottom panels}: $E_{total}=-2.12$.}
\label{Fig:12}
\end{figure*}
The motion of a test particle in the gravitational field of a massive circle and of the central mass may occur in closed trajectories as shown in Fig.~\ref{Fig:10}. Two types of these trajectories entwine both the central mass and the MC (Fig.~\ref{Fig:10}~a,b). The "eight" \, shape trajectory (Fig.\ref{Fig:10}~a) is similar to a closed trajectory in the gravitational field of just the ring \citep{Bliokh2003}. In the case of the MC alone,  without central mass, the particle will move in a "eight" \,  trajectory around the MC with an intersection point  coincident with the center of symmetry of the MC. A more complex closed trajectory is shown in Fig.~\ref{Fig:10}~b. We will conditionally call this type of orbit a "fish" \, trajectory.  
Another closed trajectory with "trefoil" shape, located near  the MC, is 
shown in Fig.~\ref{Fig:10}~c. Numerical simulations prove that these orbits 
are stable for a long time interval. Such trajectories exist only for 
this kind of axisymmetric gravitational field where the forces act in 
opposite directions. Since, with increasing distance from the ring, the potential of such a system tends to that of a point mass, the orbit of the particle tends to a circular (Fig.~\ref{Fig:10}~d). Closed compound trajectories can tell us about the possible existence of the third integral in such a potential (see, for example, \citep{Henon1964, Binney1987}).

The closed orbits are the special cases. The most common ones are regular orbits (Fig.~\ref{Fig:11}) that are typical for an axisymmetric potential. The bounded regular trajectories around the central mass and the MC are shown in Fig.~\ref{Fig:11}~a,b. The diffuse eight-type orbit (Fig.~\ref{Fig:11}~a) is obtained by the initial conditions which are slightly different from those in Fig.~\ref{Fig:10}~a. The increasing of the initial coordinate of the particle leads to a change in the shape of the trajectory that has the diffuse "dovetail" type of  orbit (Fig.~\ref{Fig:11}~b). There is also another type of regular orbits ("cup" type) which encircle only the central mass (Fig.~\ref{Fig:11}~c). If the initial value of the $\zeta$- coordinate is decreased, then the cup rotates by an angle $\pi/2$ (Fig.~\ref{Fig:11}~d). Indeed, in this case there is the essential influence of gravitational forces from the equatorial  plane of the system which explains the rotation of the orbit. These trajectories can represent  a motion in the meridional plane  as well as the projection of 3D-trajectories.

In such a system there is also a chaotic motion, the investigation of which 
requires additional analysis. Note that the appearing of chaos  for the 
relativistic system  "black hole with discs or rings"  was investigated 
with the help of the Poincar\'e section method by \citet{Semerak2010, Semerak2012}.  We consider only briefly the cases which are peculiar to 
the case under study. Fig.~\ref{Fig:12}~({\it top}) shows the Poincar\'e 
section in the phase plane ($\rho$, $V_\rho$) for a given value of the total 
energy $E_{total}=-2.17606$. Five trajectories on these maps are marked by different colors which correspond to different initial radial coordinates. The radial component of velocity, $V_{\rho}$, is obtained from the total energy law. One can see that there are four invariant points in the middle of the four loops ({\it blue points} in Fig.~\ref{Fig:12}~top) which represent stable closed orbit. For the chosen initial conditions, these points correspond to the eight-type trajectory (Fig.~\ref{Fig:10}~a). There are also three closed curves around each stable invariant points ({\it orange, green, red} curves at Fig.~\ref{Fig:12}~ {\it top}). These curves correspond to three regular trajectories similar to the diffuse eight-type orbits (Fig.~\ref{Fig:11}~a). Decreasing of the radial coordinate leads to a chaotic trajectory around the central mass and the MC. This trajectory is represented by the random distribution of points in the bounded region. The shape of the loops at the Poincar\'e section is changed for higher values of the total energy $E_{total}=-2.12$ (Fig.~\ref{Fig:12}~{\it bottom}). These loops correspond to the diffuse dovetail trajectories as in Fig.~\ref{Fig:11}~b. For further increase of the total energy,  no regular trajectories exist and only a chaotic motion is possible.  

\section{DISCUSSION}
\label{sec:discussion}

\begin{figure}
\includegraphics[width = 70mm]{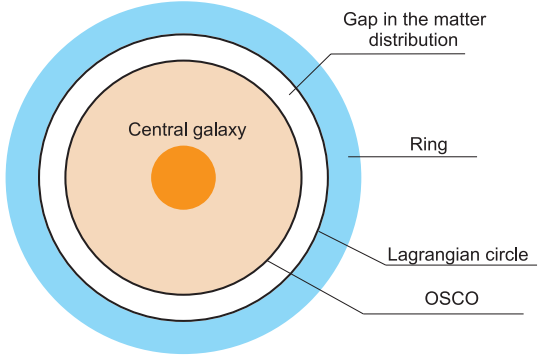}
\caption {Scheme of Hoag-like ring galaxy, with the gap in the matter distribution encircled between the outermost stable circular orbit (OSCO) and the Lagrangian circle (LC).}
\label{Fig:13}
\end{figure}

The region where the circular stable orbits do not exist may be of some interest for understanding the observational properties of those ring galaxies in which the mass of the central core is comparable with that of the ring. For instance, we can estimate the width of this region for Hoag's Object.  In this object the ring of hot, blue stars is located around the central elliptical galaxy. Using the image provided by the Hubble Heritage program a rough estimate shows that the inner edge of the ring is approximately a distance of 0.9 of the average radius of the ring itself. According to the observational result from \citep{Finkelman2011}, the mass ratio  $M_0/M_{ring} \approx 3$ and the major radius of the ring $R \approx 17''$. We can obtain the radius of the LC from (\ref{eq:17}), $\rho_{L} \approx 0.92$, which coincides with the inner edge of the ring. From equation (\ref{eq:22}), the radius of the OSCO is $\rho_{osco} \approx 0.73$. Then the width of the region from OSCO to LC is $\triangle \rho=\rho_{L}-\rho_{osco} = 0.19 R \approx 3.2''$ or, since $1''=851$~pc \citep{Springob2005}, we obtain $\triangle \rho \approx 2.8$~kpc. The scheme of a ring galaxy for the case of Hoag's Object is shown in Fig.~\ref{Fig:13} keeping corresponding proportions between main scales. The width of the region of unstable circular orbits corresponds to the  minimum observed in the average of luminosity profile for Hoag's Object \citep{Finkelman2011}. Also the minimum in the stellar distribution is seen in the  Hubble Heritage colour image of Hoag's Object.
The width of the region of unstable circular orbits is approximately the same for the core-to-ring mass ratio ($q = 0.5 \div 3$), and  it shifts to the system center if the mass of the ring increases (see Fig.~\ref{Fig:8}). This may be an additional criterion for determining the relative mass of the ring. There is some analogy with the region of unstable orbits in the case of gravitational resonance (Kirkwood gaps in the main asteroid belt, which occurs due to Jupiter's influence). In system, considered here, the appearance of the region of unstable circular orbits has an understandable physical explanation: it is a region in which the gravitational forces from the central core compete with the forces from the gravitating ring. This leads to the impossibility for the particles to move in the stable circular orbits. In this region, interactions among stars will perturb their orbits and, as a result of an unstable motion, stars can go either to the central mass or to the ring. This process can lead to the formation of a gap.  Such a gap in the distribution of stars should be inherent in any ring galaxy, in which the mass of the ring is comparable with the mass of the central galaxy. 

It was discussed in \citep{Bannikova2012} that, since the maximum of the torus potential is shifted relatively to its  cross-section center, the gravitating torus must be compressed along the major radius. In order to compensate for this compression, centrifugal forces are needed, and, hence, the orbital motion of matter is important. Therefore, the central mass plays an important role for the torus stability. For the torus with a mass of about few percent of the central mass, the orbital motion is sufficient to stabilize it. This is the case, for example, of a dusty torus in an active galactic nucleus. However, if the mass of the torus (ring) is comparable with the central mass, the torus will be compressed along the minor radius due to self-gravity effects. In order to compensate for this compression, it is necessary to apply centrifugal forces over the minor radius. It means that the motion along the minor radius could be  needed for the stabilization of a heavy torus. Perhaps  some balance is realized between the orbital motion of the ring and the vortex motion of its substance over a minor radius in the ring galaxies, which leads to the stability of such systems. 

\section{CONCLUSION}
\label{sec:conclusion}

We have analyzed the particle motion in the gravitational field of a torus with a central mass, replacing the torus potential by the potential of a massive circle. This simplification is possible because  the outer potential of the torus matches quite accurately that of a massive circle with the same mass. This allowed us to obtain the analytical solution for the radius of the circle, which we call "Lagrangian circle" (LC), where the forces by the massive ring and the central core balance each other. We show that in the meridional plane of the ring there are closed and regular trajectories of different kinds.  Investigation of the dynamics  in  the equatorial plane of such a system  shows that  there is a last stable circular orbit related to the disappearance of the extrema of the effective potential. The stable circular motion is possible only inside this last stable circular orbit, which we call "the outermost stable circular orbit" (OSCO) in analogy with a relativistic case. In the relativistic case there is the innermost stable circular orbit (ISCO) and the stable circular motion exists outside the ISCO. The OSCO can be a useful tool for investigating Hoag-like galaxies as well as ISCO is used in order to model black holes via the surrounding accretion disc. In the system considered here, there is a region between OSCO and LC where the circular motion is impossible. Such a region can lead to the formation of a stellar matter gap in Hoag-like ring systems where the central mass is comparable with the mass of the ring. 

\section*{Acknowledgments}

I thank Massimo Capaccioli and Victor Kontorovich for very fruitful discussions and useful remarks.  I am grateful to Alexey Moiseev for the discussion of the observational data for Hoag's Object. I also like to thank the anonymous referees for helpful comments and very good ideas that improved the manuscript.

\bibliographystyle{mnras} 
\bibliography{BannikovaRevised} 

\begin{thebibliography}{}
\makeatletter
\relax
\def\mn@urlcharsother{\let\do\@makeother \do\$\do\&\do\#\do\^\do\_\do\%\do\~}
\def\mn@doi{\begingroup\mn@urlcharsother \@ifnextchar [ {\mn@doi@}
  {\mn@doi@[]}}
\def\mn@doi@[#1]#2{\def\@tempa{#1}\ifx\@tempa\@empty \href
  {http://dx.doi.org/#2} {doi:#2}\else \href {http://dx.doi.org/#2} {#1}\fi
  \endgroup}
\def\mn@eprint#1#2{\mn@eprint@#1:#2::\@nil}
\def\mn@eprint@arXiv#1{\href {http://arxiv.org/abs/#1} {{\tt arXiv:#1}}}
\def\mn@eprint@dblp#1{\href {http://dblp.uni-trier.de/rec/bibtex/#1.xml}
  {dblp:#1}}
\def\mn@eprint@#1:#2:#3:#4\@nil{\def\@tempa {#1}\def\@tempb {#2}\def\@tempc
  {#3}\ifx \@tempc \@empty \let \@tempc \@tempb \let \@tempb \@tempa \fi \ifx
  \@tempb \@empty \def\@tempb {arXiv}\fi \@ifundefined
  {mn@eprint@\@tempb}{\@tempb:\@tempc}{\expandafter \expandafter \csname
  mn@eprint@\@tempb\endcsname \expandafter{\@tempc}}}

\bibitem[\protect\citeauthoryear{{Abramowicz} \& {Fragile}}{{Abramowicz} \&
  {Fragile}}{2013}]{Abramowicz2013}
{Abramowicz} M.~A.,  {Fragile} P.~C.,  2013, \mn@doi [Living Reviews in
  Relativity] {10.12942/lrr-2013-1}, \href
  {http://adsabs.harvard.edu/abs/2013LRR....16....1A} {16, 1}

\bibitem[\protect\citeauthoryear{{Antonucci}}{{Antonucci}}{1993}]{Antonucci1993}
{Antonucci} R.,  1993, \mn@doi [\araa] {10.1146/annurev.aa.31.090193.002353},
  \href {http://adsabs.harvard.edu/abs/1993ARA%26A..31..473A} {31, 473}

\bibitem[\protect\citeauthoryear{{Bannikova}, {Vakulik}  \&
  {Shulga}}{{Bannikova} et~al.}{2011}]{Bannikova2011}
{Bannikova} E.~Y.,  {Vakulik} V.~G.,   {Shulga} V.~M.,  2011, \mn@doi [\mnras]
  {10.1111/j.1365-2966.2010.17700.x}, \href
  {http://adsabs.harvard.edu/abs/2011MNRAS.411..557B} {411, 557}

\bibitem[\protect\citeauthoryear{{Bannikova}, {Vakulik}  \&
  {Sergeev}}{{Bannikova} et~al.}{2012}]{Bannikova2012}
{Bannikova} E.~Y.,  {Vakulik} V.~G.,   {Sergeev} A.~V.,  2012, \mn@doi [\mnras]
  {10.1111/j.1365-2966.2012.21186.x}, \href
  {http://adsabs.harvard.edu/abs/2012MNRAS.424..820B} {424, 820}

\bibitem[\protect\citeauthoryear{{Binney} \& {Tremaine}}{{Binney} \&
  {Tremaine}}{1987}]{Binney1987}
{Binney} J.,  {Tremaine} S.,  1987, {Galactic dynamics}

\bibitem[\protect\citeauthoryear{{Bliokh} \& {Kontorovich}}{{Bliokh} \&
  {Kontorovich}}{2003}]{Bliokh2003}
{Bliokh} K.~Y.,  {Kontorovich} V.~M.,  2003, \mn@doi [Soviet Journal of
  Experimental and Theoretical Physics] {10.1134/1.1591211}, \href
  {http://adsabs.harvard.edu/abs/2003JETP...96..985B} {96, 985}

\bibitem[\protect\citeauthoryear{{Brosch}, {Finkelman}, {Oosterloo}, {Jozsa}
  \& {Moiseev}}{{Brosch} et~al.}{2013}]{Brosch2013}
{Brosch} N.,  {Finkelman} I.,  {Oosterloo} T.,  {Jozsa} G.,   {Moiseev} A.,
  2013, \mn@doi [\mnras] {10.1093/mnras/stt1348}, \href
  {http://adsabs.harvard.edu/abs/2013MNRAS.435..475B} {435, 475}

\bibitem[\protect\citeauthoryear{{Finkelman} \& {Brosch}}{{Finkelman} \&
  {Brosch}}{2011}]{FinkelmanBrosch2011}
{Finkelman} I.,  {Brosch} N.,  2011, \mn@doi [\mnras]
  {10.1111/j.1365-2966.2011.18330.x}, \href
  {http://adsabs.harvard.edu/abs/2011MNRAS.413.2621F} {413, 2621}

\bibitem[\protect\citeauthoryear{{Finkelman}, {Moiseev}, {Brosch}  \&
  {Katkov}}{{Finkelman} et~al.}{2011}]{Finkelman2011}
{Finkelman} I.,  {Moiseev} A.,  {Brosch} N.,   {Katkov} I.,  2011, \mn@doi
  [\mnras] {10.1111/j.1365-2966.2011.19601.x}, \href
  {http://adsabs.harvard.edu/abs/2011MNRAS.418.1834F} {418, 1834}

\bibitem[\protect\citeauthoryear{{Fukushima}}{{Fukushima}}{2016}]{Fukushima2016}
{Fukushima} T.,  2016, \mn@doi [\aj] {10.3847/0004-6256/152/2/35}, \href
  {http://adsabs.harvard.edu/abs/2016AJ....152...35F} {152, 35}

\bibitem[\protect\citeauthoryear{{Gradshteyn}, {Ryzhik}, {Jeffrey}  \&
  {Zwillinger}}{{Gradshteyn} et~al.}{2007}]{Gradshteyn2007}
{Gradshteyn} I.~S.,  {Ryzhik} I.~M.,  {Jeffrey} A.,   {Zwillinger} D.,  2007,
  {Table of Integrals, Series, and Products}

\bibitem[\protect\citeauthoryear{{Henon} \& {Heiles}}{{Henon} \&
  {Heiles}}{1964}]{Henon1964}
{Henon} M.,  {Heiles} C.,  1964, \mn@doi [\aj] {10.1086/109234}, \href
  {http://adsabs.harvard.edu/abs/1964AJ.....69...73H} {69, 73}

\bibitem[\protect\citeauthoryear{{Hoag}}{{Hoag}}{1950}]{Hoag1950}
{Hoag} A.~A.,  1950, \mn@doi [\aj] {10.1086/106427}, \href
  {http://adsabs.harvard.edu/abs/1950AJ.....55Q.170H} {55, 170}

\bibitem[\protect\citeauthoryear{{Hur{\'e}} \& {Pierens}}{{Hur{\'e}} \&
  {Pierens}}{2009}]{Hure2009}
{Hur{\'e}} J.-M.,  {Pierens} A.,  2009, \mn@doi [\aap]
  {10.1051/0004-6361/200912348}, \href
  {http://adsabs.harvard.edu/abs/2009A%26A...507..573H} {507, 573}

\bibitem[\protect\citeauthoryear{{Jaffe} et~al.,}{{Jaffe}
  et~al.}{2004}]{Jaffe2004}
{Jaffe} W.,  et~al., 2004, \mn@doi [\nat] {10.1038/nature02531}, \href
  {http://adsabs.harvard.edu/abs/2004Natur.429...47J} {429, 47}

\bibitem[\protect\citeauthoryear{{Jee} et~al.,}{{Jee} et~al.}{2007}]{Jee2007}
{Jee} M.~J.,  et~al., 2007, \mn@doi [\apj] {10.1086/517498}, \href
  {http://adsabs.harvard.edu/abs/2007ApJ...661..728J} {661, 728}

\bibitem[\protect\citeauthoryear{{Landau} \& {Lifshitz}}{{Landau} \&
  {Lifshitz}}{1975}]{Landaufshitz1975}
{Landau} L.~D.,  {Lifshitz} E.~M.,  1975, {The classical theory of fields}

\bibitem[\protect\citeauthoryear{{Moiseev}, {Smirnova}, {Smirnova}  \&
  {Reshetnikov}}{{Moiseev} et~al.}{2011}]{Moiseev2011}
{Moiseev} A.~V.,  {Smirnova} K.~I.,  {Smirnova} A.~A.,   {Reshetnikov} V.~P.,
  2011, \mn@doi [\mnras] {10.1111/j.1365-2966.2011.19479.x}, \href
  {http://adsabs.harvard.edu/abs/2011MNRAS.418..244M} {418, 244}

\bibitem[\protect\citeauthoryear{{Moiseev}, {Khoperskov}, {Khoperskov},
  {Smirnova}, {Smirnova}, {Saburova}  \& {Reshetnikov}}{{Moiseev}
  et~al.}{2015}]{Moiseev2015}
{Moiseev} A.,  {Khoperskov} S.,  {Khoperskov} A.,  {Smirnova} K.,  {Smirnova}
  A.,  {Saburova} A.,   {Reshetnikov} V.,  2015, Baltic Astronomy, \href
  {http://adsabs.harvard.edu/abs/2015BaltA..24...76M} {24, 76}

\bibitem[\protect\citeauthoryear{{Mutlu Pakdil}, {Mangedarage}, {Seigar}  \&
  {Treuthardt}}{{Mutlu Pakdil} et~al.}{2017}]{MutluPakdil2017}
{Mutlu Pakdil} B.,  {Mangedarage} M.,  {Seigar} M.~S.,   {Treuthardt} P.,
  2017, \mn@doi [\mnras] {10.1093/mnras/stw3107}, \href
  {http://adsabs.harvard.edu/abs/2017MNRAS.466..355M} {466, 355}

\bibitem[\protect\citeauthoryear{{Ninkovic} \& {Jovanovic}}{{Ninkovic} \&
  {Jovanovic}}{2009}]{Ninkovic2009}
{Ninkovic} S.,  {Jovanovic} B.,  2009, \mn@doi [Serbian Astronomical Journal]
  {10.2298/SAJ0978029N}, \href
  {http://adsabs.harvard.edu/abs/2009SerAJ.178...29N} {178, 29}

\bibitem[\protect\citeauthoryear{{Schweizer}, {Ford}, {Jedrzejewski}  \&
  {Giovanelli}}{{Schweizer} et~al.}{1987}]{Schweizer1987}
{Schweizer} F.,  {Ford} Jr. W.~K.,  {Jedrzejewski} R.,   {Giovanelli} R.,
  1987, \mn@doi [\apj] {10.1086/165562}, \href
  {http://adsabs.harvard.edu/abs/1987ApJ...320..454S} {320, 454}

\bibitem[\protect\citeauthoryear{{Semer{\'a}k} \& {Sukov{\'a}}}{{Semer{\'a}k}
  \& {Sukov{\'a}}}{2010}]{Semerak2010}
{Semer{\'a}k} O.,  {Sukov{\'a}} P.,  2010, \mn@doi [\mnras]
  {10.1111/j.1365-2966.2009.16003.x}, \href
  {http://adsabs.harvard.edu/abs/2010MNRAS.404..545S} {404, 545}

\bibitem[\protect\citeauthoryear{{Semer{\'a}k} \& {Sukov{\'a}}}{{Semer{\'a}k}
  \& {Sukov{\'a}}}{2012}]{Semerak2012}
{Semer{\'a}k} O.,  {Sukov{\'a}} P.,  2012, \mn@doi [\mnras]
  {10.1111/j.1365-2966.2012.21630.x}, \href
  {http://adsabs.harvard.edu/abs/2012MNRAS.425.2455S} {425, 2455}

\bibitem[\protect\citeauthoryear{{Springob}, {Haynes}, {Giovanelli}  \&
  {Kent}}{{Springob} et~al.}{2005}]{Springob2005}
{Springob} C.~M.,  {Haynes} M.~P.,  {Giovanelli} R.,   {Kent} B.~R.,  2005,
  \mn@doi [\apjs] {10.1086/431550}, \href
  {http://adsabs.harvard.edu/abs/2005ApJS..160..149S} {160, 149}

\bibitem[\protect\citeauthoryear{{Tsupko}, {Bisnovatyi-Kogan}  \&
  {Jefremov}}{{Tsupko} et~al.}{2016}]{Tsupko2016}
{Tsupko} O.~Y.,  {Bisnovatyi-Kogan} G.~S.,   {Jefremov} P.~I.,  2016, \mn@doi
  [Gravitation and Cosmology] {10.1134/S0202289316020158}, \href
  {http://adsabs.harvard.edu/abs/2016GrCo...22..138T} {22, 138}

\bibitem[\protect\citeauthoryear{{Whitmore}, {Lucas}, {McElroy},
  {Steiman-Cameron}, {Sackett}  \& {Olling}}{{Whitmore}
  et~al.}{1990}]{Whitmore1990}
{Whitmore} B.~C.,  {Lucas} R.~A.,  {McElroy} D.~B.,  {Steiman-Cameron} T.~Y.,
  {Sackett} P.~D.,   {Olling} R.~P.,  1990, \mn@doi [\aj] {10.1086/115614},
  \href {http://adsabs.harvard.edu/abs/1990AJ....100.1489W} {100, 1489}

\bibitem[\protect\citeauthoryear{{Woodward}, {Sankaran}  \&
  {Tohline}}{{Woodward} et~al.}{1992}]{Woodward1992}
{Woodward} J.~W.,  {Sankaran} S.,   {Tohline} J.~E.,  1992, \mn@doi [\apj]
  {10.1086/171576}, \href {http://adsabs.harvard.edu/abs/1992ApJ...394..248W}
  {394, 248}

\makeatother
\end{thebibliography}

\end{document}